\documentclass[
floatfix,
twocolumn,
showpacs,
preprintnumbers,
amsmath,amssymb,pra,superscriptaddress,longbibliography]{revtex4-2}
\usepackage{graphicx}% Include figure files
\usepackage{dcolumn}% Align table columns on decimal point
\usepackage{bm}% bold math
\usepackage{color}
\usepackage{xcolor}
\usepackage{braket}
\usepackage{amsmath, amssymb}
\usepackage{booktabs}
\usepackage{tabularx}
\usepackage{hyperref}

\newcommand{\bfr}{\bm{r}}
\newcommand{\bR}{\bm{R}}
\newcommand{\ubr}{\underline{\bm{r}}}
\newcommand{\ubR}{\underline{\bm{R}}}

\newcommand{\s}{_\mathrm{{\scriptscriptstyle S}}}
\newcommand{\h}{_\mathrm{{\scriptscriptstyle H}}}
\newcommand{\xc}{_\mathrm{{\scriptscriptstyle XC}}}

\newcommand{\bk}{\bm{k}}

\begin{document}

\title{Predicting electronic structures at any length scale with machine learning}
%\title{Electronic structure prediction at any length scale with machine learning}

\author{Lenz Fiedler}
\affiliation{Center for Advanced Systems Understanding, Helmholtz-Zentrum Dresden-Rossendorf, Untermarkt 20, G\"orlitz, 02826, Saxony, Germany}

\author{Normand Modine}
\affiliation{Computational Materials and Data Science, Sandia National Laboratories, 1515 Eubank Blvd, Albuquerque, 87123, NM, USA}

\author{Steve Schmerler}
\affiliation{Information Services and Computing, Helmholtz-Zentrum Dresden-Rossendorf, Bautzner Landstraße 400, Dresden, 01328, Saxony, Germany}

\author{Dayton J. Vogel}
\affiliation{Computational Materials and Data Science, Sandia National Laboratories, 1515 Eubank Blvd, Albuquerque, 87123, NM, USA}

\author{Gabriel A. Popoola}
\affiliation{Elder Research, Inc., 300 West Main Street, Charlottesville, 22903, VA, USA}

\author{Aidan Thompson}
\affiliation{Center for Computing Research, Sandia National Laboratories 1515 Eubank Blvd, Albuquerque, 87123, NM USA}

\author{Sivasankaran Rajamanickam}
\affiliation{Center for Computing Research, Sandia National Laboratories 1515 Eubank Blvd, Albuquerque, 87123, NM USA}
\email{srajama@sandia.gov}

\author{Attila Cangi}
\affiliation{Center for Advanced Systems Understanding, Helmholtz-Zentrum Dresden-Rossendorf, Untermarkt 20, G\"orlitz, 02826, Saxony, Germany}
\email{a.cangi@hzdr.de}

\date{\today}

\maketitle

% Sort of like an abstract, but not counted towards word limit. Will be used verbatim as abstract in the online version. Max word count is 200. It requires a very specific structure: "It is aimed at readers outside the discipline. This summary paragraph should be structured as follows: 2-3 sentences of basic-level introduction to the field; a brief account of the background and rationale of the work; a statement of the main conclusions (introduced by the phrase 'Here we show' or its equivalent); and finally, 2-3 sentences putting the main findings into general context so it is clear how the results described in the paper have moved the field forwards." Also see the annotated example: https://www.nature.com/documents/nature-summary-paragraph.pdf

\section*{Abstract} %Word limit: 150
The properties of electrons in matter are of fundamental importance. They give rise to virtually all material properties and determine the physics at play in objects ranging from semiconductor devices to the interior of giant gas planets. Modeling and simulation of such diverse applications rely primarily on density functional theory (DFT), which has become the principal method for predicting the electronic structure of matter. While DFT calculations have proven to be very useful, their computational scaling limits them to small systems. We have developed a machine learning framework for predicting the electronic structure on any length scale. It shows up to three orders of magnitude speedup on systems where DFT is tractable and, more importantly, enables predictions on scales where DFT calculations are infeasible. Our work demonstrates how machine learning circumvents a long-standing computational bottleneck and advances materials science to frontiers intractable with any current solutions.

\section*{Introduction}
Electrons are elementary particles of fundamental importance. Their quantum mechanical interactions with each other and with atomic nuclei give rise to the plethora of phenomena we observe in chemistry and materials science. Knowing the probability distribution of electrons in molecules and materials $-$ their electronic structure $-$ provides insights into the reactivity of molecules, the structure and the energy transport inside planets, and how materials break.
Hence, both an understanding and the ability to manipulate the electronic structure in a material propels novel technologies impacting both industry and society. In light of the global challenges related to climate change, green energy, and energy efficiency, the most notable applications that require an explicit insight into the electronic structure of matter include the search for better batteries\cite{KaMe2006,LuLe2016} and the identification of more efficient catalysts\cite{ZhTr2020,HaGi2021}. The electronic structure is furthermore of great interest to fundamental physics as it determines the Hamiltonian of an interacting many-body quantum system\cite{HoKo1964} and is observable using experimental techniques\cite{Na2011}.

The quest for predicting the electronic structure of matter dates back to Thomas\cite{Th1927}, Fermi\cite{Fe1926}, and Dirac\cite{Di1930} who formulated the very first theory in terms of electron density distributions. While computationally cheap, their theory was not useful for chemistry or materials science due to its lack of accuracy, as pointed out by Teller\cite{Te1962}. Subsequently, based on a mathematical existence proof\cite{HoKo1964}, the seminal work of Kohn and Sham\cite{KoSh1965} provided a smart reformulation of the electronic structure problem in terms of modern density functional theory (DFT) that has led to a paradigm shift.
Due to the balance of accuracy and computational cost it offers, DFT has revolutionized chemistry $-$ with the Nobel Prize in 1998 to Kohn\cite{Ko1999} and Pople\cite{Po1999} marking its breakthrough. It is the reason DFT remains by far the most widely used method for computing the electronic structure of matter. With the advent of exascale high-performance computing systems, DFT continues reshaping computational materials science at an even bigger scale\cite{Jo2015,PaJa2019}. However, even with an exascale system, the scale one could achieve with DFT is limited due its cubic scaling on system size. We address this limitation and demonstrate that an approach based on machine learning can predict electronic structures at any length scale for the first time.

In principle, DFT is an exact method, even though in practice the exchange-correlation functional needs to be approximated\cite{LeBi2016}. Sufficiently accurate approximations do exist for useful applications, and the search for ever more accurate functionals that extend the scope of DFT is an active area of research\cite{MeBu2017} where methods of artificial intelligence and machine learning (ML) have led to great advances in accuracy\cite{KiMc2021,PeKa2022} without addressing the scaling limitation.

Despite these initial successes, DFT calculations are hampered inherently due to their computational cost. The standard algorithm scales as the cube of system size, limiting routine calculations to problems comprised of only a few hundred atoms.
This is a fundamental limitation that has impeded large-scale computational studies in chemistry and materials science so far.
Lifting the curse of cubic scaling has been a long-standing challenge.
Prior works have attempted to overcome this challenge in terms of either an orbital-free formulation of DFT \cite{LiCa2005} or algorithmic development known as linear-scaling DFT \cite{Ya1991,GoCo1994}. Neither of these paths has led to a general solution to this problem. 
More recently, other works have explored leveraging ML techniques to circumvent the inherent bottleneck of the DFT algorithm. These have used kernel-ridge regression \cite{BrVo2017} or neural networks \cite{TsMi2020, MiRy2019}, but remained on the conceptual level and are applicable to only model systems, small molecules, and low-dimensional solids.

Despite all these efforts, computing the electronic structure of matter at large scales while maintaining first-principles accuracy has remained an elusive goal so far. We provide a solution to this long-standing challenge in the form of a linear-scaling ML surrogate for DFT. Our algorithm enables accurate predictions of the electronic structure of materials at any length scale.

\section*{Results}
\subsection*{Ultra-large scale electronic structure predictions with neural networks}
In this work, we circumvent the computational bottleneck of DFT calculations by utilizing neural networks in local atomic environments to predict the local electronic structure. Thereby, we achieve the ability to compute the electronic structure of matter at any length scale with minimal computational effort and at the first-principles accuracy of DFT.

To this end, we train a feed-forward neural network $M$ that performs a simple mapping
\begin{equation}
    \tilde{d}(\epsilon, \bfr) = M(B(J, \bfr)) \; , \label{eq:NNPassing}
\end{equation}
where the bispectrum coefficients $B$ of order $J$ serve as \textit{descriptors} that encode the positions of atoms relative to every point in real space $\bfr$, while $\tilde{d}$ approximates the local density of states (LDOS) $d$ at energy $\epsilon$. The LDOS encodes the local electronic structure at each point in real space and energy. More specifically, the LDOS can be used to calculate the electronic density $n$ and density of states $D$, two important quantities which enable access to a range of observables such as the total free energy $A$ itself \cite{ellis2021accelerating}, i.e.,
\begin{equation}
    A[n,D] = A\Big{[}n[d],D[d]\Big{]} = A [d] \; .
\end{equation}

The key point is that the neural network is trained locally on a given point in real space and therefore has no awareness of the system size. Our underlying working assumption relies on the nearsightedness of the electronic structure \cite{Ko1996}. It sets a characteristic length scale beyond which effects on the electronic structure decay rapidly with distance. Since the mapping defined in Eq.~(\ref{eq:NNPassing}) is purely local, i.e., performed individually for each point in real space, the resulting workflow is scalable across the real-space grid, highly parallel, and transferable to different system sizes. Non-locality is factored into the model via the bispectrum descriptors, which are calculated by including information from adjacent points in space up to a specified cutoff radius consistent with the aforementioned length scale. 

\begin{figure*}[htp]
    \centering
    \includegraphics[width=0.85\textwidth]{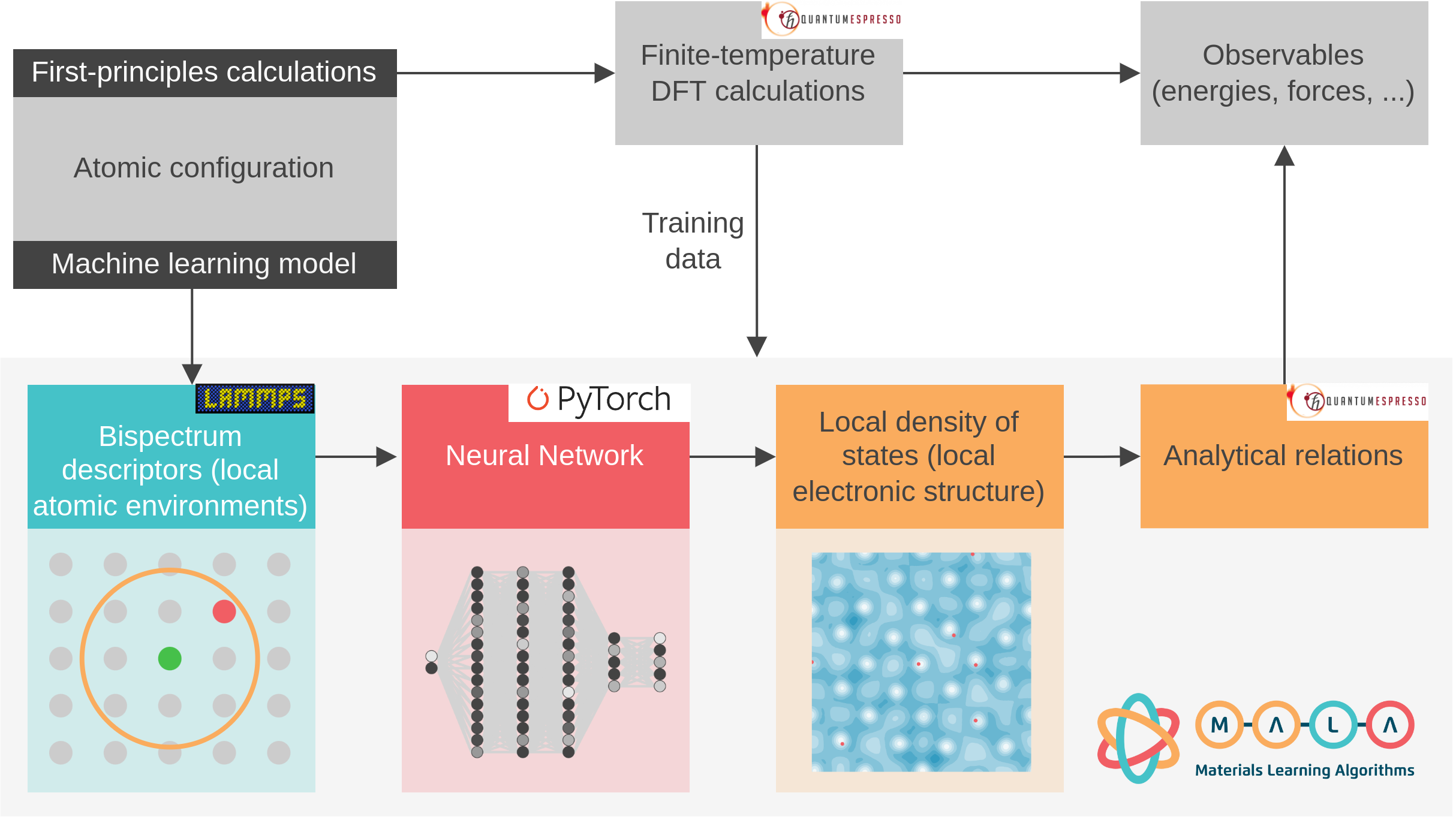}
    \caption{Overview of the MALA framework. ML models created via this workflow can be trained on data from popular first-principles simulation codes such as Quantum ESPRESSO \cite{giannozzi_quantum_2009}. The pictograms below the individual workflow steps show, from left to right, the calculation of local descriptors at an arbitrary grid point (green) based on information at adjunct grid points (grey) within a certain cutoff radius (orange), with an atom shown in red; a neural network; the electronic structure, exemplified here as a contour plot of the electronic density for a cell containing Aluminum atoms (red).} 
    %The pictograms are adapted from Ref.~\cite{fiedler2022electronic}.}
    \label{fig:malaworkflow}
\end{figure*}

The individual steps of our computational workflow are visualized in Fig.~\ref{fig:malaworkflow}. They include combining the calculation of bispectrum descriptors to encode the atomic density, training and evaluation of neural networks to predict the LDOS, and finally, the post-processing of the LDOS to physical observables. The entire workflow is implemented end-to-end as a software package called Materials Learning Algorithms (MALA)~\cite{Cangi_mala_2021}, where we employ interfaces to popular open-source software packages, namely LAMMPS~\cite{LAMMPS} (descriptor calculation), PyTorch~\cite{paszke_pytorch_2019} (neural network training and inference), and Quantum ESPRESSO~\cite{giannozzi_quantum_2009} (post-processing of the electronic structure data to observables). 

\begin{figure*}[htp]
    \centering
    \includegraphics[width=0.85\textwidth]{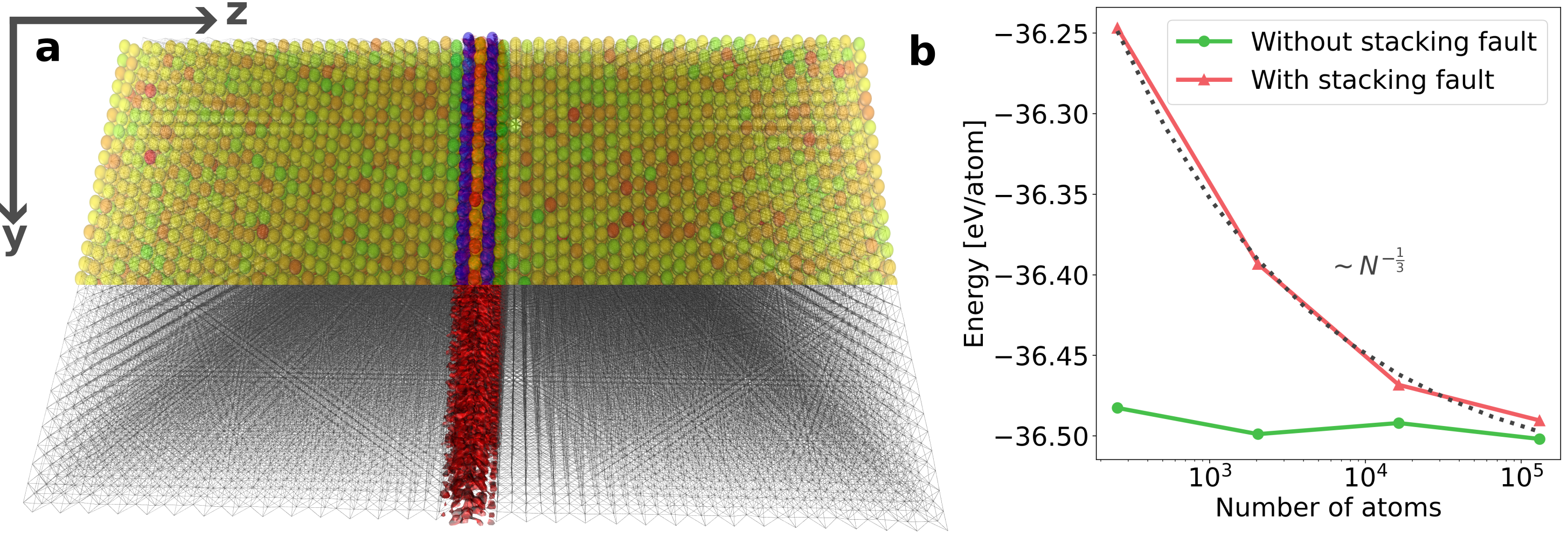}
    \caption{Illustrating size transferability of our ML model. \textbf{a}: Beryllium simulation cell of 131,072 atoms with a stacking fault, generated by shifting three layers along the y-axis creating a local fcc geometry, as opposed to the hcp crystal structure of Beryllium. The colors in the upper half correspond to the centrosymmetry parameter calculated by OVITO \cite{OVITO}, where blue corresponds to fcc and red-to-light-green to hcp local geometries. The lower half of the image, generated with VMD \cite{vmd}, shows the difference in the electronic density for 131,072 Beryllium atoms with and without a stacking fault. \textbf{b}: Energy differences due to introducing a stacking fault into Beryllium cells of differing sizes. }
    \label{fig:large_density}
\end{figure*}

We illustrate our workflow by computing the electronic structure of sample material that contains more than 100,000 atoms. The employed ML model is a feed-forward neural network that is trained on simulation cells containing 256 Beryllium atoms. In Fig.~\ref{fig:large_density}, we showcase how our framework predicts multiple observables at previously unattainable scales. 
Here, we show an atomic snapshot containing 131,072 Beryllium atoms at room temperature into which a stacking fault has been introduced, i.e., three atomic layers have been shifted laterally, changing the local crystal structure from hcp to fcc. Our ML model is then used to predict both the electronic densities and energies of this simulation cell with and without the stacking fault. As expected, our ML predictions reflect the changes in the electronic density due to the changes in the atomic geometry. 
The energetic differences associated with such a stacking fault are expected to follow a behavior $\sim N^{-\frac{1}{3}}$, where $N$ is the number of atoms. By calculating the energy of progressively larger systems with and without a stacking fault, we find that this expected behavior is indeed obeyed quite closely by our model (Fig.~\ref{fig:large_density}\textbf{b}).

Our findings open up the possibility to train models for specific applications on scales previously unattainable with traditional simulation methods. Our ML predictions on the 131,072 atom system take 48 minutes on 150 standard CPUs; the resulting computational cost of roughly 121 CPU hours (CPUh) is comparable to a conventional DFT calculation for a few hundred atoms.
The computational cost of our ML workflow is orders of magnitude below currently existing linear-scaling DFT codes, i.e., codes scaling with $\sim N$ \cite{nakata_large_2020}, which employ approximations in terms of the density matrix. Their computational cost lies two orders of magnitude above our approach. Standard DFT codes scale even more unfavorably as $\sim N^3$, which renders simulations like the one presented here completely infeasible.

Common research directions for utilizing ML in the realm of electronic structure theory either focus on predicting energies and forces of extended systems (ML interatomic potentials\cite{wood_data-driven_2019}) or directly predicting observables of interest such as polarizabilities\cite{wilkins_accurate_2019}. MALA models are not limited to singular observables and even give insight into the electronic structure itself, from which a range of relevant observables including the total free energy, the density of states, the electronic density, and atomic forces follow. 

The utility of our ML framework for chemistry and materials science relies on two key aspects. It needs to scale well with system size up to the 100,000 atom scale and beyond. Furthermore, it also needs to maintain accuracy as we run inferences on increasingly large systems. Both issues are addressed in the following.

\subsection*{Computational scaling}
\begin{figure*}[htp]
    \centering
    \includegraphics[width=0.85\textwidth]{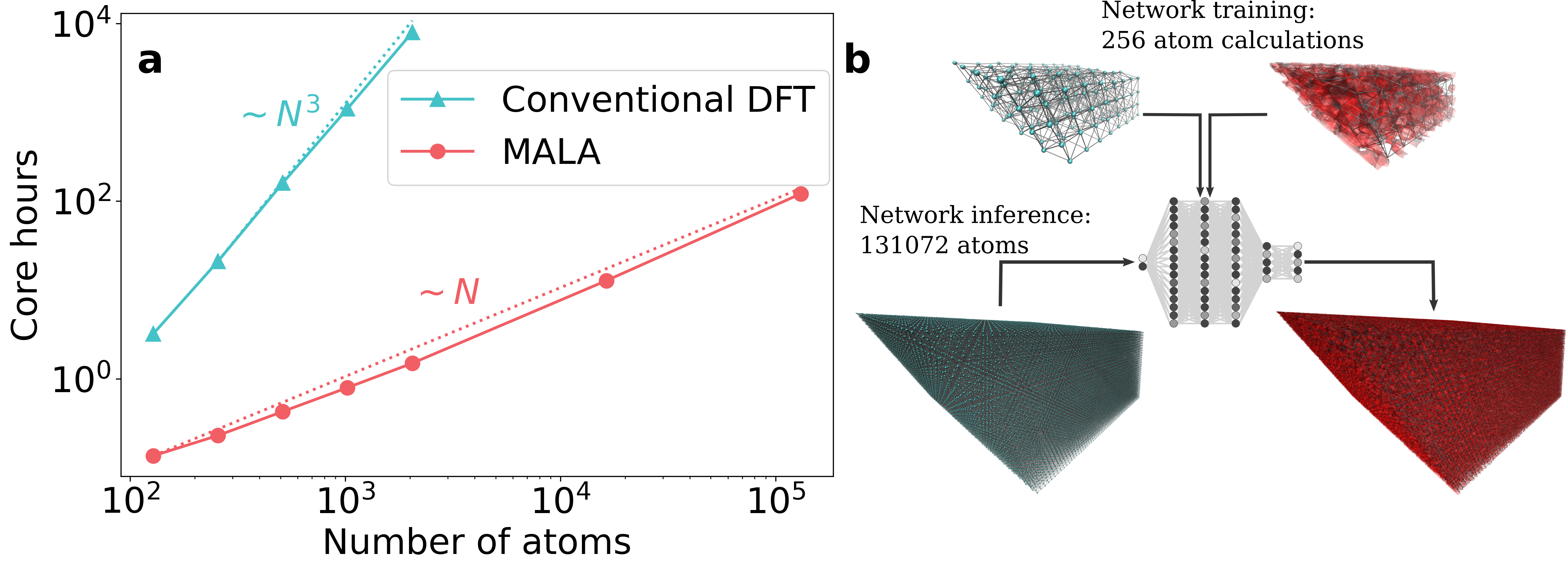}
    \caption{Scaling behavior of MALA. \textbf{a}: Comparison of the scaling behavior of conventional DFT (QuantumESPRESSO) and the MALA framework with the number of atoms. Please note that for the sake of consistency, slightly different computational parameters have been used for the DFT calculations here compared to the DFT reference calculations in Fig.~\ref{fig:size_transfer_accuracy_superplot}. \textbf{b}: General workflow of size transferability in MALA. }
    \label{fig:scaling_plot}
\end{figure*}

The computational cost of conventional DFT calculations scales as $N^3$. Improved algorithms can enable an effective $N^2$ scaling in certain cases over certain size ranges\cite{kresse_ab_1993}. In either case, one is faced with an increasingly insurmountable computational cost for systems involving more than a few thousand atoms. As illustrated in Fig.~\ref{fig:scaling_plot}\textbf{a}, conventional DFT calculations (here using the Quantum ESPRESSO\cite{giannozzi_quantum_2009} software package) are subject to this scaling behavior. 
Contrarily, the computational cost of using MALA models for size extrapolation (as shown in Fig.~\ref{fig:scaling_plot}\textbf{b}) grows linearly with the number of atoms and has a significantly smaller computational overhead. We observe speed-ups of up to three orders of magnitude for atom counts up to which DFT calculations are computationally tractable.   

MALA model inference consists of three steps. First, the descriptor vectors are calculated on a real-space grid, then the LDOS is computed using a pre-trained neural network for given input descriptors, and finally, the LDOS is post-processed to compute electronic densities, total energies, and other observables. The first two parts of this workflow trivially scale with $N$, since they strictly perform operations per grid point, and the real space simulation grid grows linearly with $N$. 

Obtaining linear scaling for the last part of the workflow, which includes processing the electronic density to the total free energy, is less trivial since it requires both the evaluation of the ion-ion energy as well as the exchange-correlation energy, which for the pseudopotentials we employ includes the calculation of non-linear core corrections. While both of these terms can be shown to scale linearly with system size in principle, in practice this requires the addition of a few custom routines, as is further outlined in the methods section.

\subsection*{Accuracy and transferability to large scales}
When assessing the transferability of our workflow, we are faced with the problem that we cannot compute conventional DFT results beyond about 2,000 atoms due to the high cost of these calculations. We, therefore, split the task of evaluating the predictive performance into first showing that our model retains its competitive accuracy when comparing predictions with DFT reference data above the training data size, and thereafter asserting that this trend holds when going to ultra-large scales of hundreds of thousands of atoms.

\subsubsection*{Benchmarks at DFT scales ($\sim 10^3$ atoms)}
We tackle the first part of this problem by investigating a system of Beryllium atoms at room temperature and ambient mass density (1.896 g/cc). Neural networks are trained on LDOS data generated for 256 atoms. After training, inference was performed for an increasing number of atoms, namely 256, 512, 1,024, and 2,048 atoms. 

\begin{figure*}[htp]
    \centering
    \includegraphics[width=0.85\textwidth]{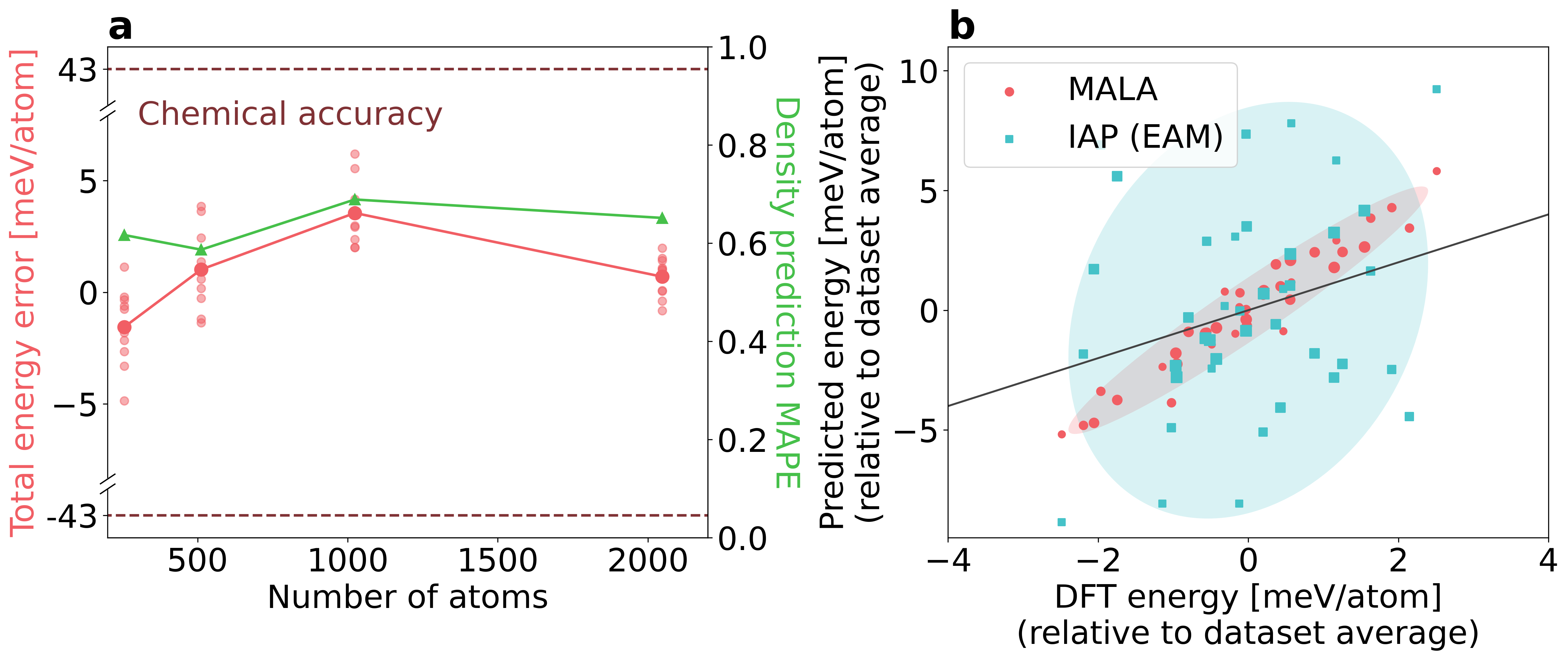}
    \caption{Accuracy of MALA. \textbf{a}: Prediction errors when using MALA to calculate total free energies and electronic densities compared to DFT data. \textbf{b}: Correlation between DFT and predicted total energies for MALA and an EAM type interatomic potential (IAP) for Beryllium (across all system sizes).}
    \label{fig:size_transfer_accuracy_superplot}
\end{figure*}

The total free energy and the electronic density were used to assess the accuracy of MALA predictions for a total of 10 atomic configurations per system size. In Fig.~\ref{fig:size_transfer_accuracy_superplot}\textbf{a} we report the absolute error of the energy and the mean absolute percentage error (MAPE) of the density across system sizes. It is evident that the errors stay roughly constant across system size and are well within both chemical accuracy (below 43 meV/atom). Furthermore, the error of the energy is within the 10 meV/atom threshold which is considered the gold standard for ML interatomic potentials. Likewise, the error in the electronic density is remarkably low, lying well under 1\%.

The accuracy of absolute total free energy predictions does not suffice to assess model performance, since one is usually interested in energy differences. Therefore, we relate the predicted total free energy to the DFT reference data set in Fig.~\ref{fig:size_transfer_accuracy_superplot}\textbf{b}. The data points are drawn across all system sizes but are given relative to the respective means per system size for the sake of readability. Ideally, the resulting distribution would lie along a straight line. In practice, both a certain spread around this line (unsystematic errors) and a tilt of the line (systematic errors) can be expected. We quantify our results by comparing MALA (red circles) with an embedded-atom-method (EAM) interatomic potential (blue squares)\cite{daw_embedded-atom_1984, BeEAM} which is commonly used in molecular dynamics simulations. It can clearly be seen that MALA outperforms the EAM model in both unsystematic as well as systematic errors, and, therefore, delivers physically correct energies beyond the system sizes it was trained on.

\subsubsection*{Accuracy at ultra-large scales ($\sim 10^5$ atoms)}
Finally, we tackle the second step of providing evidence that MALA predictions on the ultra-large scale are expected to be as accurate as conventional DFT calculations. This analysis is grounded in the local nature of our workflow. Given that the local environments are similar to those observed in training, predictions for arbitrarily large cells boil down to interpolation, a task at which neural networks excel. Accordingly, our ML model performs a perceived size extrapolation by actually performing local interpolations.

\begin{figure*}[htp]
    \centering
    \includegraphics[width=\textwidth]{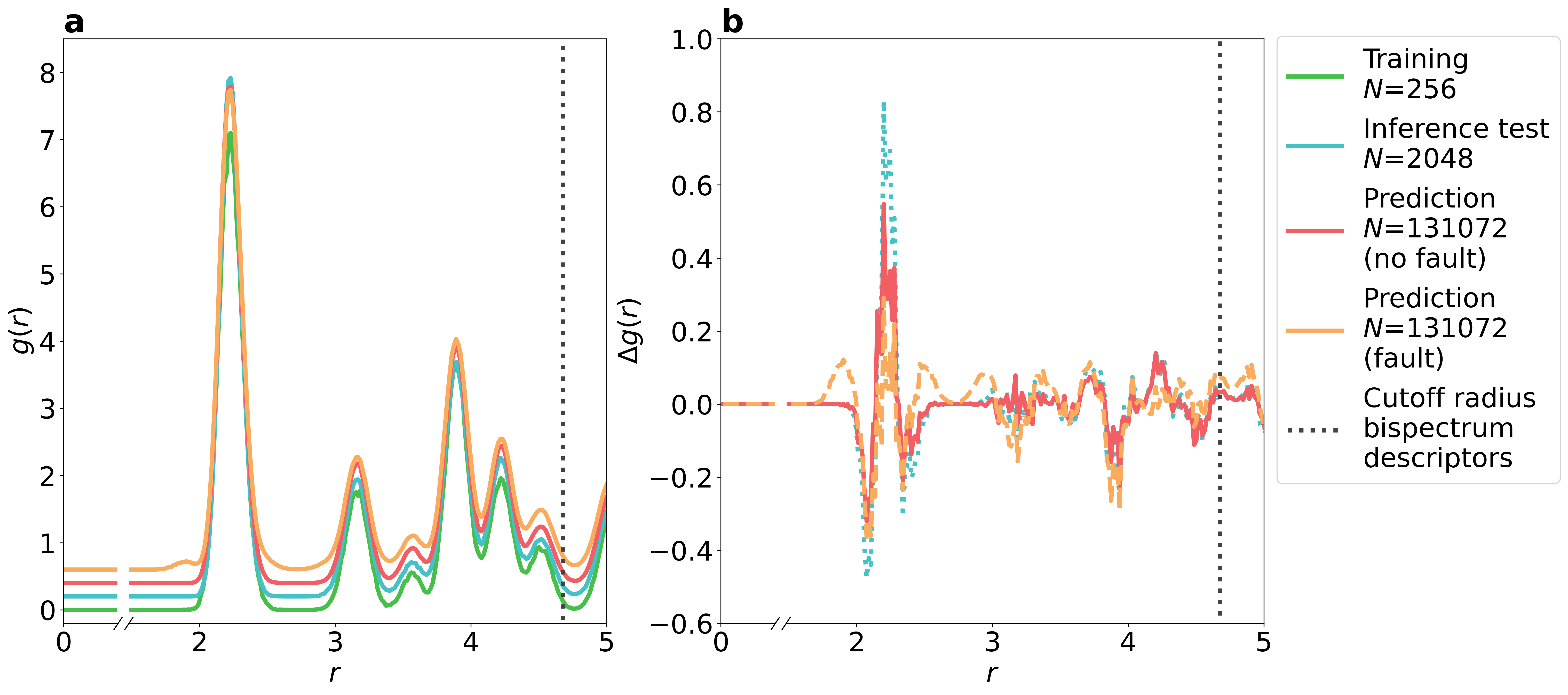}
    \caption{Analysis of size transferability. \textbf{a}: Radial distribution functions for Beryllium simulation cells of differing sizes, within the radius in which information is incorporated into bispectrum descriptors. For technical details on the radial distribution function, see the methods section.  Note that the curves of the inference test (blue) and prediction (red, orange) data sets have been shifted along the y-axis by a constant value of 0.2 to better illustrate how similar they are. \textbf{b}: Absolute difference $\Delta g(r)$ of the radial distribution functions with respect to the training data set.}
    \label{fig:statistical_anaylsis}
\end{figure*}

To verify the similarity of the atomic configurations in the training set with those used for inference at ultra-large scales, we employ the radial distribution function. It is a useful quantity that distinguishes between different phases of a material, by giving insight into how likely it is to find an atom at a given distance from a reference point. 
Since the input to our workflow, $B$, is calculated based on atomic densities drawn from a certain cutoff radius, a matching radial distribution function up to this point indicates that the individual input vectors $B$ should on average be similar between simulation cells. This comparison is shown in Fig.~\ref{fig:statistical_anaylsis} where the radial distribution functions $g(r)$ of the training (256 atoms, green), inference test (2,048 atoms, blue), and ultra-large prediction (131,072 atoms, red and orange) data sets are illustrated. Fig.~\ref{fig:statistical_anaylsis}\textbf{a} illustrates the absolute values, whereas Fig.~\ref{fig:statistical_anaylsis}\textbf{b} shows the difference of the radial distribution function to the training data set. It should be noted that for the sake of comparability, the radial distribution functions for 256 atoms and 2,048 were averaged over 30 atomic configurations and 10 atomic configurations, respectively. Furthermore, we do not show the full radial distribution function for radii below 1.5 \AA, since it is zero irrespective of the number of atoms, due to the average interatomic distance for this system. Note also that we have shifted these curves along the y-axis by a constant value of 0.2 from each other to better illustrate how similar they are.

In Fig.~\ref{fig:statistical_anaylsis} slight deviations between the radial distribution functions of different system sizes can be seen, most notably for the cells containing a stacking fault. Overall, these deviations are small in magnitude, especially for the unperturbed cells, and generally, all radial distribution functions agree very well up to the cutoff radius (dotted black) from which information is incorporated into the bispectrum descriptors.

This analysis hence provides evidence that training, inference test, and the ultra-large simulation cells possess, on average, the same local environments. It indicates that our MALA predictions of the electronic structure and energy are based on interpolations on observed data. Therefore our models can be expected to be accurate at ultra-large scales far exceeding those for which reference data exists. By comparing the radial distribution functions for 2,048 and 131,072 atoms, we can deduce that errors similar to those reported in Fig.~\ref{fig:size_transfer_accuracy_superplot} can be assumed for these ultra-large scales.

\section*{Discussion}
We have introduced an ML model that avoids the computational bottleneck of DFT calculations. It scales linearly with system size as opposed to conventional DFT that follows a cubic scaling. Our ML model enables efficient electronic structure predictions at scales far beyond what is tractable with conventional DFT, in fact at any length scale. In contrast to existing ML approaches, our workflow provides direct access to the electronic structure and is not limited to specific observables. Any physical quantity that can be expressed as a functional of the electronic density, the fundamental quantity in DFT, can be predicted using the ML models trained with the workflow presented here.

At system sizes where DFT benchmarks are still available, we demonstrate that our ML model is capable of reproducing energies and electronic densities of extended systems at virtually no loss in accuracy while outperforming other ML models that are based solely on energy. Furthermore, we demonstrate that our ML workflow enables predicting the electronic structure for systems with more than 100,000 atoms at a very low computational cost. We underpin its accuracy at these ultra-large scales by analyzing the radial distribution function and find that our ML models can be expected to deliver accurate results even at such length scales.

We expect our ML model to set new standards in a number of ways. 
Using our ML model either directly or in conjunction with other ML workflows, such as ML interatomic potentials for pre-sampling of atomic configurations, will enable first-principles modeling of materials without finite-size errors. Combined with Monte-Carlo sampling and atomic forces from automatic differentiation, our ML model can replace ML interatomic potentials and yield thermodynamic observables at much higher accuracy. Another application our ML model enables is the prediction of electronic densities in semiconductor devices, for which an accurate modeling capability at the device scale has been notoriously lacking.   
Finally, we also expect our ML model to pave the way to predicting electronic phase transitions on a quantitative level as it resolves changes in the electronic structure at hitherto unattainable length scales.

% \subfile{methods}

% \bibliography{mala_size_transfer.bib}

% \bibliographystyle{Science}

\newpage

\section*{Methods}

\makeatother
\renewcommand{\tablename}{Tab.}

\label{sec:methods}
\subsection*{Density Functional Theory}
Density functional theory (DFT) is the most widely used method for computing (thermodynamic) materials properties in chemistry and materials science because it strikes a balance between computational cost and accuracy. Within DFT, one commonly seeks to describe a coupled system of $N$ ions of charge $Z_\alpha$ at collective positions $\ubR=(\bR_1, \bR_2, ..., \bR_N)$ and $L$ electrons at collective positions $\ubr=(\bfr_1, \bfr_2, ..., \bfr_L)$ on a quantum statistical mechanical level~\cite{HoKo1964,mermin_thermal_1965}. Within the commonly assumed Born-Oppenheimer approximation~\cite{born_zur_1927}, the Hamiltonian 
\begin{equation}
    \hat{H} =\hat{T}+\hat{V}^{ee}+\hat{V}^{ei} \label{eq:Hamiltonian_full} \; ,
\end{equation}
represents a system of interacting electrons in the external field of the ions that are simplified to classical point particles. Here, $\hat{T}=\sum_{j}^{L} -\nabla_j^2/2$ denotes the kinetic energy operator of the electrons, $\hat{V}^{ee}=\sum_{i}^{L} \sum_{j \ne i}^{L} 1/(2\vert\bfr_i-\bfr_j\vert)$ the electron-electron interaction, and $\hat{V}^{ei}= -\sum_{j}^{L} \sum_{\alpha}^{N} Z_\alpha/\vert\bfr_j-\bR_\alpha\vert$ the external potential, i.e., the electron-ion interaction. The Born-Oppenheimer Hamiltonian separates the electronic and ionic problems into a quantum mechanical and classical mechanical problem. Such an assumption is feasible since ionic masses far exceed the electronic mass, leading to vastly different time scales for movement and equilibration.

At finite temperatures $\tau>0$, the theoretical description is extended to the grand canonical operator~\cite{toda1983statistical} 
\begin{equation}
    \hat{\Omega} = \hat{H} - \tau \hat{S}-\mu \hat{N}\label{eq:GrandCanonicalOperator} \; ,
\end{equation}
where $\hat{S}=-k_B\ln\hat{\Gamma}$ denotes the entropy operator, $\hat{N}$ the particle-number operator, and $\mu$ the chemical potential. Here, we introduced the statistical density operator $\hat{\Gamma} = \sum_{L,m}w_{L,m} \vert\Psi_{L,m}\rangle \langle\Psi_{L,m}\vert$ with the $L$-electron eigenstates $\Psi_{L,m}$ of the Hamiltonian $\hat{H}$ and $w_{L,m}$ as the normalized statical weights that obey $\sum_{L,m} w_{L,m}=1$. Any observable $A$ is then computed as an average
\begin{equation}
    A[\hat{\Gamma}] = \mathrm{Tr}\{\hat{\Gamma}\hat{A}\}
    = \sum_{L,m} w_{L,m} \langle\Psi_{L,m}\vert\hat{A}\vert\Psi_{L,m}\rangle\ .
\end{equation}
Most importantly, finding the grand potential
\begin{equation}
    \Omega[\hat{\Gamma}] = \mathrm{Tr}\{\hat{\Gamma}\hat{\Omega}\} \; ,
    \label{eq:GrandCanonicalPotential}
\end{equation}
amounts to finding a $\hat{\Gamma}$ that minimizes this expression.
The exact solution to this problem evades numerical treatment even with modern hardware and software due to the electron-electron interaction in the Born-Oppenheimer Hamiltonian of Eq.~(\ref{eq:Hamiltonian_full}). It dictates an exponential growth of complexity with the number of electrons $L$, i.e., $~e^{L}$.

Based on the theorems of Hohenberg and Kohn~\cite{HoKo1964} and of Mermin~\cite{mermin_thermal_1965}, DFT makes solving this problem computationally tractable by employing the electronic density $n$ as the central quantity. The formal scaling reduces to $L^3$ due to the Kohn-Sham approach~\cite{KoSh1965}. Within DFT, all quantities of interest are formally defined as functionals of the electronic density via a one-to-one correspondence with the external (here, electron-ion) potential. In conjunction with the Kohn-Sham scheme, which introduces an auxiliary system of non-interacting fermions restricted to reproduce the density of the interacting system practical calculations become feasible. Rather than evaluating Eq.~(\ref{eq:GrandCanonicalPotential}) using many-body wave functions $\Psi_{L}$, the grand potential is evaluated as a functional of density $n$ as
\begin{align}
   \Omega[n] = &T\s[n] - k_\mathrm{B}\tau S\s[n] \nonumber+ E\h[n] \\ &+ E\xc[n] + E^{ei}[n] - \mu L \label{eq:GrandCanonicalPotential_KS} \; ,
\end{align}
with the kinetic energy of the Kohn-Sham system $T\s$, the entropy of the Kohn-Sham system $S\s$, the classical electrostatic interaction energy $E\h$ (Hartree energy), the electrostatic interaction energy of the electronic density with the ions $V^{ei}$, and the exchange-correlation (free) energy $E\xc$. The Kohn-Sham system serves as an auxiliary system that is used to calculate the kinetic energy and entropy terms in Eq.~(\ref{eq:GrandCanonicalPotential_KS}). The Kohn-Sham equations are defined as a system of one-electron Schrödinger-like equations
\begin{equation}
	\left[-\frac{1}{2}\nabla^2 + v\s(\bfr)\right]\phi_j(\bfr) = \epsilon_j \phi_j(\bfr) \label{eq:KSequation.nonzerotemp} \; ,
\end{equation}
with an effective potential, the Kohn-Sham potential $v\s(\bfr)$,
that yields the electronic density of the interacting system via 
\begin{equation}
    n(\bfr) = \sum_j f^\tau(\epsilon_j)\, \mid\phi_j(\bfr)\mid^2 \; ,
\end{equation}
where $\phi_j$ denotes the Kohn-Sham orbitals, $\epsilon_j$ the Kohn-Sham eigenalues, and $f^\tau(\epsilon_j)$ the Fermi-Dirac distribution at temperature $\tau$. The Kohn-Sham potential is a single-particle potential defined as $v\s(\bfr) = v^{ei}(\bfr) + v\h(\bfr) + v\xc(\bfr)$, where  $v^{ei}(\bfr)=-\sum_{\alpha}^{N} Z_\alpha/\vert\bfr-\bR_\alpha\vert$ denotes the electron-ion interaction potential, $v\h[n](\bfr) = \delta E\h[n]/\delta n(\bfr) = \int d\bfr'\, n(\bfr')/|\bfr-\bfr'|$ the Hartree potential, and $v\xc[n](\bfr)=\delta E\xc[n]/\delta n(\bfr)$ the exchange-correlation potential. Note that within the Kohn-Sham framework at finite temperatures, several quantities including $v\s(\bfr)$, $\epsilon_j$, $\phi_j$, $n$, $T\s$, $S\s$, and $E\xc$ are technically temperature dependent; we omit to label this temperature dependency explicitly in the following for the sake of brevity. The Kohn-Sham formalism of DFT is formally exact if the correct form of the exchange-correlation functional $E\xc[n]$ was known. In practice, approximations of the exchange-correlation functional are employed. There exists a plethora of useful functionals both for the ground state (such as the LDA~\cite{KoSh1965, ceperley_ground_1980}, PBE~\cite{perdew_accurate_1986,perdew_accurate_1992,perdew_generalized_1996}, and SCAN~\cite{sun_strongly_2015} functionals) and at finite temperature~\cite{brown_2013_exchange,karasiev_nonempirical_2013, karasiev_2014_accurate,groth_2017_ab}. Such functionals draw on different ingredients for approximating the exchange-correlation energy. Some rely only on the electronic density (e.g.,~LDA), while others incorporate density gradients (e.g.,~PBE) or even the kinetic energy density (e.g.,~SCAN). Consequently, functionals differ in their provided accuracy and application domain like molecules or solids.

The calculation of dynamical properties is enabled in this framework via the estimation of the atomic forces, which are then used to time-propagate the ions in a process called DFT molecular dynamics (DFT-MD). The forces are evaluated via the total free energy as $-{\partial A[n](\ubR)}/{\partial \ubR}$, where the total free energy $A^\mathrm{BO}_\mathrm{total}$ is obtainedd from Eq.~(\ref{eq:GrandCanonicalPotential_KS}) as $A^\mathrm{BO}_\mathrm{total}[n](\ubR) = \Omega[n] + \mu L$.
While this framework can be employed to calculate a number of (thermodynamic) materials properties~\cite{MDreview}, the treatment of systems of more than roughly a thousand atoms becomes computationally intractable due to the $L^2$ to $L^3$ scaling typically observed when running DFT calculations for systems in this size range. Therefore, current research efforts are increasingly focused on the combination of machine learning (ML) and DFT methods~\cite{FiSh2022}.

\subsection*{DFT surrogate models}
Machine learning (ML) comprises a number of powerful algorithms, that are capable of learning, i.e., improving through data provided to them. Within DFT and computational materials science in general, ML is often applied in one of two settings, as shown in Ref.~\cite{FiSh2022}. Firstly, ML algorithms learn to predict specific properties of interest (e.g.,~structural or electronic properties) and thus bypass the need to perform first-principles simulations for investigations across vast chemical parameter spaces. Secondly, ML algorithms may provide direct access to atomic forces and energies, and thus accelerate dynamical first-principles simulations drastically, resulting in ML interatomic potentials (ML-IAPs) for MD simulations. 

We have recently introduced an ML framework that does not fall in either category, as it comprises a DFT surrogate model that replaces DFT for predicting a range of useful properties~\cite{ellis2021accelerating}. Our framework directly predicts the electronic structure of materials and is therefore not restricted to singular observables. 
Within this framework, the central variable is the local density of states (LDOS) defined by 
\begin{equation}
d(\epsilon, \bfr) = \sum_j {\mid\phi_j(\bfr)\mid}^2 \delta(\epsilon-\epsilon_j) \; . \label{eq:ldos}
\end{equation}
The merit of using the LDOS as a central variable is that it determines both the electronic density as 
\begin{align}
    n(\bfr) &= \sum_j f^\tau(\epsilon_j)\, \mid\phi_j(\bfr)\mid^2 = \int d \epsilon\;  f^\tau(\epsilon) d(\epsilon, \bfr) \; ,
\end{align}
and the density of states (DOS) as
\begin{align}
    D(\epsilon) &= \sum_j \delta(\epsilon-\epsilon_j) = \int d\bfr \; d(\epsilon, \bfr)  \; .
\end{align}
As opposed to related work~\cite{chandrasekaran_solving_2019}, we use these two quantities to calculate the total free energy drawing on a reformulation of Eq.~(\ref{eq:GrandCanonicalPotential_KS}), which expresses all energy terms dependent on the KS wave functions and eigenvalues in terms of the DOS. More precisely, by employing the band energy
\begin{equation}
    E_b = \int d \epsilon\;  f^\tau(\epsilon) \epsilon D(\epsilon) \; ,
\end{equation}
and reformulating the electronic entropy in terms of the DOS, i.e., 
\begin{align}
	S\s = &- \sum_{j} \Big[ f^\tau_j(\epsilon_j) \ln{f^\tau_j(\epsilon_j)} \nonumber\\ &+\left(1-f^\tau_j(\epsilon_j)\right)\ln \left(1-f^\tau_j(\epsilon_j)\right) \Big] \;  \nonumber\\
	= & -\int d\epsilon \; \big( f^\tau(\epsilon) \ln{\left[f^\tau(\epsilon)\right]}  \nonumber\\ &+ \left[1-f^\tau(\epsilon)\right] \ln{\left[1-f^\tau(\epsilon)\right]}\big) D(\epsilon) \; ,
\end{align}
the total free energy $A^\mathrm{BO}_\mathrm{total}$ can be expressed as 
\begin{align}
    A^\mathrm{BO}_\mathrm{total}[d] = &E_b[D[d]] - \tau S\s[D[d]] - E\h[n[d](\bfr)] \nonumber \\ &+ E\xc[n[d](\bfr)]  - \int d\bfr\, v\xc(\bfr)n[d](\bfr) \; , \label{eq:TotalFreeEnergy_ldos}
\end{align}
where $D$ and $n$ are functionals of the LDOS and $v\xc(\bfr)=\delta E\xc[n[d](\bfr)]/\delta n[d](\bfr)$. 

In our framework, the LDOS is learned \textit{locally}. For each point in real space, the respective LDOS (a vector in the energy domain) is predicted separately from adjacent points. Non-locality enters this prediction through the \textit{descriptors} that serve as input to the ML algorithm. Here, we chose bispectrum descriptors~\cite{thompson_spectral_2015} denoted as $B$. In contrast to their usual application as a basis for interatomic potentials, these descriptors are employed to encode local information on atomic neighborhoods at each point in space. This is done by evaluating the total density of neighbor atoms via a sum of delta functions
\begin{equation}
    	\rho ({\bfr}) = \delta(\bm{0}) +   \sum_{r_{k} < R_\mathrm{cut}^{\nu_k}}  {f_c(\vert\bfr_{k}\vert, R_\mathrm{cut}^{\nu_k}) w_{\nu_k} \delta({\bf r}_{k})}\,.
	\label{eq:atomic_density}
\end{equation}
In Eq.~(\ref{eq:atomic_density}), the sum is performed over all $k$ atoms within a cutoff distance $ R_\mathrm{cut}^{\nu_k}$ using a switching function $f_c$ that ensures smoothness of atomic contributions at the edges of the sphere with radius $ R_\mathrm{cut}^{\nu_k}$. These atoms are located at position $\bm{r}_k$ relative to the grid point $\bfr$, while the chemical species $\nu_k$ enters the equation via the dimensionless weights $w_{\nu_k}$. The thusly defined density is then expanded into a basis of 4D hyperspherical harmonic functions, eventually yielding the descriptors $B(J,\bfr)$ with a feature dimension $J$ (see Ref.~\cite{ellis2021accelerating},\cite{thompson_spectral_2015}). Constructing descriptors in such a way introduces two hyperparameters, $ R_\mathrm{cut}^{\nu_k}$, which determines the radius from which information is incorporated into the descriptors and $J_\mathrm{max}$, which determines the number of hyperspherical harmonics used for the expansion, i.e., the dimensionality of the descriptor vectors. As we have shown recently in Ref.~\cite{fiedler2022electronic}, they can be chosen accurately without the need for ML model training and inference based on similarity measures that agree with physical intuition.  

A mapping from $B(J,\bfr)$ to $d(\epsilon, \bfr)$ is now performed via a neural network (NN), $M$, i.e.,
\begin{equation}
    \tilde{d}(\epsilon, \bfr) = M(B(J, \bfr)) \; , \label{eq:NNPassing_supp}
\end{equation}
where $\tilde{d}$ is the approximate LDOS. After performing such a network pass for each point in space, the resulting approximate LDOS can be post-processed into the observables mentioned above. 

We employ NNs, because they are, in principle, capable of approximating any given function~\cite{hornik_approximation_1991}. In the present case, we employ feed-forward NNs~\cite{minsky_perceptrons_1987} which consist of a sequence of layers containing individual artificial neurons~\cite{rosenblatt_perceptron_1957} that are fully connected to each neuron in subsequent layers. Each layer is a transformation of the form 
\begin{equation}
    \bm{x}^{\ell + 1} = \varphi({\bf W}^{\ell} \bm{x}^{\ell} + \bm{b}^{\ell}) \; ,
\end{equation}
that maps $\bm{x}$ from layer $\ell$ to $\ell+1$ by addition of a bias vector $\bm{b}$, matrix multiplication with a weight matrix ${\bf W}$, and an \textit{activation function} $\varphi$. For the DFT surrogate models discussed here, the input to the first transformation $\bm{x}^0$ is $B(J,\bfr)$ for a specific point in space $\bfr$; the output of the last layer $\bm{x}^\mathcal{L} $ is $d(\epsilon, \bfr)$ for the same $\bfr$. The number of layers $\mathcal{L} $ and activation function $\varphi$ have to be determined through prior \textit{hyperparameter optimization}, among other \textit{hyperparameters} such as the width of the individual layers. In Ref.~\cite{fiedler2022electronic} we show how such a hyperparameter optimization can be drastically improved upon in terms of computational effort, while the hyperparameters employed for this study are detailed in Sec.~\ref{sec:computational_details}. For each architecture of the NN, the weights and biases have to be optimized using gradient-based updates in a process called \textit{training} based on a technique called \textit{backpropagation}~\cite{rumelhart1986learning} which is carried out using gradients averaged over portions of the data (so-called \textit{mini-batches}); other technical parameters include stopping criteria for the \textit{early stopping} of the model optimization and the \textit{learning rate} for the gradient-based updates~\cite{deeplearningbook}.

Unlike DFT, the great majority of operations in our DFT surrogate model have a computational cost that naturally scales linearly with system size:  (1) the descriptors are evaluated independently at each point on the computational grid using algorithms in LAMMPS that take advantage of the local dependence of the descriptors on the atomic positions in order to maintain linear scaling;  (2) the NN is evaluated independently at each grid point in order to obtain the LDOS at each point;  (3) the DOS is evaluated by a reduction over grid points, the Fermi level is found, and $E_b$ and $S\s$ are evaluated;  (4) the density is calculated independently at each grid point; and (5) three-dimensional Fast Fourier transforms, which are implemented efficiently in Quantum ESPRESSO, are used to evaluate $E\h$ from the density.  The remaining terms are $E\xc$, $v\xc$, and the ion-ion interaction energy.  The exchange-correlation terms can almost be evaluated independently at each point (using Fast Fourier transforms to evaluate gradients if necessary), but the pseudo-potentials that we use include non-linear core corrections, which require the addition of a ``core density'' centered on each atom to the density used to calculate $E\xc$ and $v\xc$.  Likewise, the ion-ion interaction energy can be evaluated efficiently using Fast Fourier transforms if we can compute the sum of non-overlapping charge distributions containing the appropriate ionic charge centered on each atom.  The key to calculating these terms with a computational cost that scales linearly with system size is an efficient algorithm to evaluate the structure factor.

If $F(\bfr)$ is some periodic function represented by its values on the computational grid, its Fast Fourier transform $\tilde{F}(\bm{G})$ gives its representation in the basis of plane-waves $\exp{(i \bm{G} \cdot \bfr)}$, where the reciprocal lattice vectors $\bm{G}$ form a reciprocal-space grid with the same dimensions as the computational grid. The structure factor is defined as
\begin{equation}
    \tilde{S}(\bm{G}) = \sum_{\alpha} \exp(i \bm{G} \cdot \bR_{\alpha}) \; ,
\end{equation}
where the summation over atom positions $\bR_{\alpha}$ runs over all atoms within one copy of the periodically repeated computational cell.  The structure factor is very useful because
\begin{equation}
     F^{S}(\bfr) \equiv \sum_{\alpha} F(\bfr - \bR_{\alpha}) \; ,
\end{equation}
can be efficiently evaluated as the inverse Fourier transform of $\tilde{F}^{S} = \tilde{S}(\bm{G})\tilde{F}(\bm{G})$.  Thus, the structure factor can be used to evaluate the non-linear core correction density and the ion-ion interaction energy when evaluating the DFT total energy. However, the straightforward evaluation of $\tilde{S}(\bm{G})$ on the grid of $\bm{G}$ vectors scales as the square of the system size.  We circumvent this bottleneck by taking advantage of the real-space localization properties of the Gaussian function $G(\bfr)$ in order to efficiently evaluate $G^{S}(\bfr)$ within the LAMMPS code~\cite{LAMMPS}.  Then, within Quantum ESPRESSO, we use a fast Fourier transformation to calculate $\tilde{G}^{S}(\bm{G})$, and the structure factor is obtained as
\begin{equation}
     \tilde{S}(\bm{G}) = \frac{\tilde{G}^{S}(\bm{G})}
     {\tilde{G}(\bm{G})}\; .
\end{equation}
A suitable choice of the Gaussian width for $G(\bfr)$ allows us to minimize aliasing errors due to Fourier components beyond the Nyquist limit of the computational grid, while also maintaining good precision in the above division. 

\subsection*{Data analysis}
\label{sec:statistical_analysis}

We assess whether our models are employed in an interpolative setting when applied to larger cells. 
To that end, we analyze the radial distribution function, which is defined as
\begin{equation}
    % Lenz' eq.
    %%g(r) = \frac{\Xi(r)}{\rho N V_\mathrm{shell}} \; ,
    %
    % Most general textbook eq.
    %
    %%g(r) = \frac{V}{N^2} \left\langle \sum_{i=1}^N \sum_{\substack{j=1\\j\neq i}}^N
    %%            \delta(r - \left\vert{\bfr}_i-{\bfr}_j\right\vert)\right\rangle
    %
    g(r) = \frac{1}{\rho\,N\,V(r)}\,\sum_{i=1}^N \sum_{\substack{j=1\\j\neq i}}^N
                \delta(r - \left\vert{\bfr}_i-{\bfr}_j\right\vert)\;.
\end{equation}
It is the average ion density in a shell $[r, r+dr]$ of volume $V(r)$ around a reference ion at $r=0$, relative to an isotropic system of density $\rho=N/V$~\cite{rdf}. The radial distribution function is often used to identify different phases of a material and, in our case, it can be used to verify that simulation cells with differing numbers of atoms are equivalent in their ion distribution up to a certain cutoff radius. For technical reasons, there exists an upper radius up to which $g$ is well-defined, which is a result of the minimum image convention~\cite{metropolis1953} and which amounts to half the cell edge length in case of cubic cells. For small cells, the employed cutoff radius lies slightly beyond this radius, but this does not affect model inference, since periodic boundary conditions are applied for the calculation of the bispectrum descriptors.

\subsection*{Computational details}
\label{sec:computational_details}

\subsubsection*{Training Data}
Increasingly larger DFT-MD simulations at 298K have been performed to acquire atomic configurations for simulation cells containing 256 to 2,048 Beryllium atoms. DFT-MD calculations up to 512 atoms have been carried out using Quantum ESPRESSO, while simulations for 1,024 and 2,048 atoms have been performed using VASP~\cite{kresse_ab_1993,kresse_efficient_1996, kresse_efficiency_1996}. In either case, DFT-MD simulations have been performed at the $\Gamma$-point, using a plane-wave basis set with an energy cutoff of 40 Ry (Quantum ESPRESSO) or 248 eV (VASP), and an ultrasoft pseudopotential~\cite{QEPseudopotentials} (Quantum ESPRESSO) or a PAW pseudopotential~\cite{blochl_projector_1994,kresse_ultrasoft_1999} (VASP). The resulting trajectories have been analyzed with a method akin to the equilibration algorithm outlined in prior work~\cite{Orbital_Free_DFTMD}, although here equilibration thresholds have been defined manually. Thereafter, snapshots have been sampled from these trajectories such that the minimal euclidean distance between any two atoms within the last sampled snapshot and potentially next sampled snapshots lies above the empirically determined threshold of 0.125 \AA.  The resulting data set of Beryllium at room temperature includes ten configurations per system size, except for 256 atoms, where a larger number of configurations is needed to enable the training and verification of models. For all of these configurations, DFT calculations have been carried out with Quantum ESPRESSO, using the aforementioned cutoff and pseudopotential. The Brillouin zone has been sampled by Monkhorst-Pack~\cite{monkhorst_special_1976} sampling, with the number of $\bm{k}$-points given in Tab.~\ref{tab:k_grids}.

The employed calculation parameters have been determined via a convergence analysis with a threshold of 1 meV/atom, except for Beryllium systems with 2,048 atoms, where only $\Gamma$ point calculations have been performed due to computational constraints. 

The values in Tab.~\ref{tab:k_grids} refer to those DFT calculations that were performed to gather reference energies and densities. To calculate the LDOS, one has to employ larger $\bm{k}$-grids, as the discretiatzion of $\bm{k}$-space with a finite number of points in $\bm{k}$-space can introduce errors and features into the (L)DOS that are unphysical. As discussed in prior work~\cite{ellis2021accelerating,fiedler2022electronic}, such features can be removed by employing a larger number of $\bm{k}$-points than for typical DFT simulations. The correct $\bm{k}$-grid has to be determined through a convergence test such that no unphysical oscillations occur in the (L)DOS. By applying an established analysis~\cite{ellis2021accelerating,fiedler2022electronic}, we have determined $12\times6\times6$ as a suitable $\bm{k}$-grid for 256 Beryllium atoms. Again, Monkhorst-Pack sampling has been used.

\begin{table}[htp]
    \centering  
    \begin{tabularx}{0.4\textwidth}{ll}%cccccccc}
    	\toprule
    	\textbf{Number of atoms} &  $\bm{k}$\textbf{-grid} \ \\\midrule
    	 256 & $8\times4\times4$ \\
        512 &  $4\times4\times2$ \\
        1024 & $3\times3\times3$ \\
        2048 & $\Gamma$-point \\
        \bottomrule
    \end{tabularx}
    \caption{Overview over the $\bm{k}$-grids used for the various DFT calculations.}\label{tab:k_grids}		
\end{table}

In order to assess the scaling of DFT for Fig.~\ref{fig:scaling_plot}\textbf{a} of the main manuscript, we kept a constant $\bk$-grid were possible, in comparison to the adapted $\bk$-grids used for the reference data calculation used for Fig.~\ref{fig:size_transfer_accuracy_superplot}. More specifically, in order to reflect realistic simulation settings, we employed a $3\times3\times3$ grid, i.e., a $\bk$-grid consistent with 1,024 atoms, the largest number of atoms for which $\bk$-point converged simulations could be performed. The same number of $\bk$-points was used for 128, 256, and 512 atoms. Performing a DFT calculation for 2,048 atoms was impossible due to the large memory demand and the computational resources available. We, therefore, performed a singular 2,048 atoms calculation with a $4\times2\times2$ $\bk$-grid, utilizing more $\bk$-points in the $x$-direction, since the 2,048 atom cells are extended in that direction compared to the 1,024 atom cells. Overall, this change in $\bk$-grid leads to only a small deviation of the observed $\sim N^3$ behavior.

\subsubsection*{Machine Learning Models}

For all ML experiments, the architecture and hyperparameters discussed in the original MALA publication~\cite{ellis2021accelerating} have been employed. One training and one validation snapshot have been used in the 256-atom case. 

\section*{Data Availability}
Training data of the Beryllium system is publicly available~\cite{fiedler_ldossnap_2022}. Please note that this published data set~\cite{fiedler_ldossnap_2022} includes a larger data set of the Beryllium system as it has been used in multiple publications. In this manuscript, only a subset of Beryllium configurations has been used (256 atoms: 0-30; 512 atoms: 5-14; 1024 atoms: 0-9; 2048 atoms: 0-9).

\section*{Code Availability}
All calculations described within this work have been carried out with the freely available MALA code~\cite{Cangi_mala_2021} version 1.1.0. Benchmark models of the Beryllium system are publicly available~\cite{fiedler_ldossnap_2022}, as are the corresponding input scripts~\cite{sizetransferscripts}.

\section*{Acknowledgements}
The authors are grateful to the Center for Information Services and High Performance Computing [Zentrum für Informationsdienste und Hochleistungsrechnen (ZIH)] at TU Dresden for providing its facilities for high throughput calculations. We also gratefully acknowledge Alexander Debus for providing a CPU allocation on the taurus HPC system of ZIH at TU Dresden.

\section*{Author contributions} L.F.~performed all Beryllium-related calculations (DFT-MD, DFT, and MALA), code integration into the MALA code, and data visualization. N.M.~ and D.V.~ implemented the parallelization of the total energy evaluation, and N.M. eliminated scaling bottlenecks in the total energy evaluation. S.S.~carried out the extrapolation transferability analysis. A.T.~developed the parallelization of the descriptor calculation. S.R. and A.C. contributed to the theory and development of the MALA framework, supported data analysis, and supervised the overall project. All authors contributed to writing the manuscript.

\section*{Funding}
Sandia National Laboratories is a multimission laboratory managed and operated by National Technology \& Engineering Solutions of Sandia, LLC, a wholly-owned subsidiary of Honeywell International Inc., for the U.S. Department of Energy’s National Nuclear Security Administration under contract DE-NA0003525. This paper describes objective technical results and analysis. Any subjective views or opinions that might be expressed in the paper do not necessarily represent the views of the U.S. Department of Energy or the United States Government. 

This work was in part supported by the Center for Advanced Systems Understanding (CASUS) which is financed by Germany’s Federal Ministry of Education and Research (BMBF) and by the Saxon state government out of the State budget approved by the Saxon State Parliament. 

\section*{Competing interests} 
There are no competing interests.


\begin{thebibliography}{10}
	\expandafter\ifx\csname url\endcsname\relax
	\def\url#1{\texttt{#1}}\fi
	\expandafter\ifx\csname urlprefix\endcsname\relax\def\urlprefix{URL }\fi
	\providecommand{\bibinfo}[2]{#2}
	\providecommand{\eprint}[2][]{\url{#2}}
	
	\bibitem{KaMe2006}
	\bibinfo{author}{Kang, K.}, \bibinfo{author}{Meng, Y.~S.},
	\bibinfo{author}{Bréger, J.}, \bibinfo{author}{Grey, C.~P.} \&
	\bibinfo{author}{Ceder, G.}
	\newblock \bibinfo{title}{{Electrodes with High Power and High Capacity for
			Rechargeable Lithium Batteries}}.
	\newblock \emph{\bibinfo{journal}{\emph{\emph{Science}}}}
	\textbf{\bibinfo{volume}{311}}, \bibinfo{pages}{977--980}
	(\bibinfo{year}{2006}).
	
	\bibitem{LuLe2016}
	\bibinfo{author}{Lu, J.} \emph{et~al.}
	\newblock \bibinfo{title}{{A lithium{\textendash}oxygen battery based on
			lithium superoxide}}.
	\newblock \emph{\bibinfo{journal}{\emph{\emph{Nature}}}}
	\textbf{\bibinfo{volume}{529}}, \bibinfo{pages}{377--382}
	(\bibinfo{year}{2016}).
	
	\bibitem{ZhTr2020}
	\bibinfo{author}{Zhong, M.} \emph{et~al.}
	\newblock \bibinfo{title}{{Accelerated discovery of {CO}2 electrocatalysts
			using active machine learning}}.
	\newblock \emph{\bibinfo{journal}{\emph{\emph{Nature}}}}
	\textbf{\bibinfo{volume}{581}}, \bibinfo{pages}{178--183}
	(\bibinfo{year}{2020}).
	
	\bibitem{HaGi2021}
	\bibinfo{author}{Hannagan, R.~T.} \emph{et~al.}
	\newblock \bibinfo{title}{{First-principles design of a
			single-atom{\textendash}alloy propane dehydrogenation catalyst}}.
	\newblock \emph{\bibinfo{journal}{\emph{\emph{Science}}}}
	\textbf{\bibinfo{volume}{372}}, \bibinfo{pages}{1444--1447}
	(\bibinfo{year}{2021}).
	
	\bibitem{HoKo1964}
	\bibinfo{author}{Hohenberg, P.} \& \bibinfo{author}{Kohn, W.}
	\newblock \bibinfo{title}{{Inhomogeneous electron gas}}.
	\newblock \emph{\bibinfo{journal}{\emph{\emph{Phys. Rev.}}}}
	\textbf{\bibinfo{volume}{136}}, \bibinfo{pages}{B864--B871}
	(\bibinfo{year}{1964}).
	
	\bibitem{Na2011}
	\bibinfo{author}{Nakashima, P. N.~H.}, \bibinfo{author}{Smith, A.~E.},
	\bibinfo{author}{Etheridge, J.} \& \bibinfo{author}{Muddle, B.~C.}
	\newblock \bibinfo{title}{{The Bonding Electron Density in Aluminum}}.
	\newblock \emph{\bibinfo{journal}{\emph{\emph{Science}}}}
	\textbf{\bibinfo{volume}{331}}, \bibinfo{pages}{1583--1586}
	(\bibinfo{year}{2011}).
	
	\bibitem{Th1927}
	\bibinfo{author}{Thomas, L.~H.}
	\newblock \bibinfo{title}{{The calculation of atomic fields}}.
	\newblock \emph{\bibinfo{journal}{\emph{\emph{Math. Proc. Camb. Philos.
					Soc.}}}} \textbf{\bibinfo{volume}{23}}, \bibinfo{pages}{542–548}
	(\bibinfo{year}{1927}).
	
	\bibitem{Fe1926}
	\bibinfo{author}{Fermi, E.}
	\newblock \bibinfo{title}{{Zur Quantelung des idealen einatomigen Gases}}.
	\newblock \emph{\bibinfo{journal}{\emph{\emph{Z. Physik}}}}
	\textbf{\bibinfo{volume}{36}}, \bibinfo{pages}{902--912}
	(\bibinfo{year}{1926}).
	
	\bibitem{Di1930}
	\bibinfo{author}{Dirac, P. A.~M.}
	\newblock \bibinfo{title}{{Note on Exchange Phenomena in the Thomas Atom}}.
	\newblock \emph{\bibinfo{journal}{\emph{\emph{Math. Proc. Camb. Philos.
					Soc.}}}} \textbf{\bibinfo{volume}{26}}, \bibinfo{pages}{376–385}
	(\bibinfo{year}{1930}).
	
	\bibitem{Te1962}
	\bibinfo{author}{Teller, E.}
	\newblock \bibinfo{title}{{On the Stability of Molecules in the Thomas-Fermi
			Theory}}.
	\newblock \emph{\bibinfo{journal}{\emph{\emph{Rev. Mod. Phys.}}}}
	\textbf{\bibinfo{volume}{34}}, \bibinfo{pages}{627--631}
	(\bibinfo{year}{1962}).
	
	\bibitem{KoSh1965}
	\bibinfo{author}{Kohn, W.} \& \bibinfo{author}{Sham, L.~J.}
	\newblock \bibinfo{title}{{Self-Consistent Equations Including Exchange and
			Correlation Effects}}.
	\newblock \emph{\bibinfo{journal}{\emph{\emph{Phys. Rev.}}}}
	\textbf{\bibinfo{volume}{140}}, \bibinfo{pages}{A1133--A1138}
	(\bibinfo{year}{1965}).
	
	\bibitem{Ko1999}
	\bibinfo{author}{Kohn, W.}
	\newblock \bibinfo{title}{{Nobel Lecture: Electronic structure of matter---wave
			functions and density functionals}}.
	\newblock \emph{\bibinfo{journal}{\emph{\emph{Rev. Mod. Phys.}}}}
	\textbf{\bibinfo{volume}{71}}, \bibinfo{pages}{1253--1266}
	(\bibinfo{year}{1999}).
	
	\bibitem{Po1999}
	\bibinfo{author}{Pople, J.~A.}
	\newblock \bibinfo{title}{{Nobel Lecture: Quantum chemical models}}.
	\newblock \emph{\bibinfo{journal}{\emph{\emph{Rev. Mod. Phys.}}}}
	\textbf{\bibinfo{volume}{71}}, \bibinfo{pages}{1267--1274}
	(\bibinfo{year}{1999}).
	
	\bibitem{Jo2015}
	\bibinfo{author}{Jones, R.~O.}
	\newblock \bibinfo{title}{{Density functional theory: Its origins, rise to
			prominence, and future}}.
	\newblock \emph{\bibinfo{journal}{\emph{\emph{Rev. Mod. Phys.}}}}
	\textbf{\bibinfo{volume}{87}}, \bibinfo{pages}{897--923}
	(\bibinfo{year}{2015}).
	
	\bibitem{PaJa2019}
	\bibinfo{author}{{de Pablo, J.J., \emph{et al.}}}
	\newblock \bibinfo{title}{{New frontiers for the materials genome initiative}}.
	\newblock \emph{\bibinfo{journal}{\emph{\emph{npj Comput. Mater.}}}}
	\textbf{\bibinfo{volume}{5}} (\bibinfo{year}{2019}).
	
	\bibitem{LeBi2016}
	\bibinfo{author}{Lejaeghere, K.} \emph{et~al.}
	\newblock \bibinfo{title}{{Reproducibility in density functional theory
			calculations of solids}}.
	\newblock \emph{\bibinfo{journal}{\emph{\emph{Science}}}}
	\textbf{\bibinfo{volume}{351}}, \bibinfo{pages}{aad3000}
	(\bibinfo{year}{2016}).
	
	\bibitem{MeBu2017}
	\bibinfo{author}{Medvedev, M.~G.}, \bibinfo{author}{Bushmarinov, I.~S.},
	\bibinfo{author}{Sun, J.}, \bibinfo{author}{Perdew, J.~P.} \&
	\bibinfo{author}{Lyssenko, K.~A.}
	\newblock \bibinfo{title}{{Density functional theory is straying from the path
			toward the exact functional}}.
	\newblock \emph{\bibinfo{journal}{\emph{\emph{Science}}}}
	\textbf{\bibinfo{volume}{355}}, \bibinfo{pages}{49--52}
	(\bibinfo{year}{2017}).
	
	\bibitem{KiMc2021}
	\bibinfo{author}{Kirkpatrick, J.} \emph{et~al.}
	\newblock \bibinfo{title}{{Pushing the frontiers of density functionals by
			solving the fractional electron problem}}.
	\newblock \emph{\bibinfo{journal}{\emph{\emph{Science}}}}
	\textbf{\bibinfo{volume}{374}}, \bibinfo{pages}{1385--1389}
	(\bibinfo{year}{2021}).
	
	\bibitem{PeKa2022}
	\bibinfo{author}{Pederson, R.}, \bibinfo{author}{Kalita, B.} \&
	\bibinfo{author}{Burke, K.}
	\newblock \bibinfo{title}{{Machine learning and density functional theory}}.
	\newblock \emph{\bibinfo{journal}{\emph{\emph{Nat. Rev. Phys}}}}
	\textbf{\bibinfo{volume}{4}}, \bibinfo{pages}{357--358}
	(\bibinfo{year}{2022}).
	
	\bibitem{LiCa2005}
	\bibinfo{author}{Lign{\`e}res, V.~L.} \& \bibinfo{author}{Carter, E.~A.}
	\newblock \bibinfo{title}{{An Introduction to Orbital-Free Density Functional
			Theory}}.
	\newblock \emph{In \emph{\bibinfo{booktitle}{{\emph{Handbook of Materials
						Modeling: Methods}}}}}, \bibinfo{pages}{137--148}
	(\bibinfo{publisher}{Springer Netherlands}, \bibinfo{address}{Dordrecht},
	\bibinfo{year}{2005}).
	
	\bibitem{Ya1991}
	\bibinfo{author}{Yang, W.}
	\newblock \bibinfo{title}{{Direct calculation of electron density in
			density-functional theory}}.
	\newblock \emph{\bibinfo{journal}{\emph{\emph{Phys. Rev. Lett.}}}}
	\textbf{\bibinfo{volume}{66}}, \bibinfo{pages}{1438--1441}
	(\bibinfo{year}{1991}).
	
	\bibitem{GoCo1994}
	\bibinfo{author}{Goedecker, S.} \& \bibinfo{author}{Colombo, L.}
	\newblock \bibinfo{title}{{Efficient Linear Scaling Algorithm for Tight-Binding
			Molecular Dynamics}}.
	\newblock \emph{\bibinfo{journal}{\emph{\emph{Phys. Rev. Lett.}}}}
	\textbf{\bibinfo{volume}{73}}, \bibinfo{pages}{122--125}
	(\bibinfo{year}{1994}).
	
	\bibitem{BrVo2017}
	\bibinfo{author}{Brockherde, F.} \emph{et~al.}
	\newblock \bibinfo{title}{{Bypassing the Kohn-Sham equations with machine
			learning}}.
	\newblock \emph{\bibinfo{journal}{\emph{\emph{Nat. Commun.}}}}
	\textbf{\bibinfo{volume}{8}} (\bibinfo{year}{2017}).
	
	\bibitem{TsMi2020}
	\bibinfo{author}{Tsubaki, M.} \& \bibinfo{author}{Mizoguchi, T.}
	\newblock \bibinfo{title}{{Quantum Deep Field: Data-Driven Wave Function,
			Electron Density Generation, and Atomization Energy Prediction and
			Extrapolation with Machine Learning}}.
	\newblock \emph{\bibinfo{journal}{\emph{\emph{Phys. Rev. Lett.}}}}
	\textbf{\bibinfo{volume}{125}}, \bibinfo{pages}{206401}
	(\bibinfo{year}{2020}).
	
	\bibitem{MiRy2019}
	\bibinfo{author}{Mills, K.} \emph{et~al.}
	\newblock \bibinfo{title}{Extensive deep neural networks for transferring small
		scale learning to large scale systems}.
	\newblock \emph{\bibinfo{journal}{\emph{\emph{Chem. Sci.}}}}
	\textbf{\bibinfo{volume}{10}}, \bibinfo{pages}{4129--4140}
	(\bibinfo{year}{2019}).
	
	\bibitem{ellis2021accelerating}
	\bibinfo{author}{Ellis, J.~A.} \emph{et~al.}
	\newblock \bibinfo{title}{{Accelerating finite-temperature Kohn-Sham density
			functional theory with deep neural networks}}.
	\newblock \emph{\bibinfo{journal}{\emph{\emph{Phys. Rev. B}}}}
	\textbf{\bibinfo{volume}{104}}, \bibinfo{pages}{035120}
	(\bibinfo{year}{2021}).
	
	\bibitem{Ko1996}
	\bibinfo{author}{Kohn, W.}
	\newblock \bibinfo{title}{{Density Functional and Density Matrix Method Scaling
			Linearly with the Number of Atoms}}.
	\newblock \emph{\bibinfo{journal}{\emph{\emph{Phys. Rev. Lett.}}}}
	\textbf{\bibinfo{volume}{76}}, \bibinfo{pages}{3168--3171}
	(\bibinfo{year}{1996}).
	
	\bibitem{Cangi_mala_2021}
	\bibinfo{author}{Cangi, A.} \emph{et~al.}
	\newblock \bibinfo{title}{{MALA}}.
	\newblock \emph{\bibinfo{journal}{\emph{Zenodo,
				https://doi.org/10.5281/zenodo.5557254}}}  (\bibinfo{year}{2021}).
	
	\bibitem{LAMMPS}
	\bibinfo{author}{Thompson, A.~P.} \emph{et~al.}
	\newblock \bibinfo{title}{{{LAMMPS} - a flexible simulation tool for
			particle-based materials modeling at the atomic, meso, and continuum
			scales}}.
	\newblock \emph{\bibinfo{journal}{\emph{\emph{Comp. Phys. Comm.}}}}
	\textbf{\bibinfo{volume}{271}}, \bibinfo{pages}{108171}
	(\bibinfo{year}{2022}).
	
	\bibitem{paszke_pytorch_2019}
	\bibinfo{author}{Paszke, A.} \emph{et~al.}
	\newblock \bibinfo{title}{{{{{PyTorch}}: An {{Imperative Style}},
				{{High}}-{{Performance Deep Learning Library}}}}}.
	\newblock \emph{In \emph{\bibinfo{booktitle}{{\emph{Advances in {{Neural
								Information Processing Systems}}}}}}}, vol.~\bibinfo{volume}{32}
	(\bibinfo{publisher}{{Curran Associates, Inc.}},
	\bibinfo{address}{Vancouver}, \bibinfo{year}{2019}).
	
	\bibitem{giannozzi_quantum_2009}
	\bibinfo{author}{Giannozzi, P.} \emph{et~al.}
	\newblock \bibinfo{title}{{{QUANTUM ESPRESSO}}: A modular and open-source
		software project for quantum simulations of materials}.
	\newblock \emph{\bibinfo{journal}{\emph{\emph{J. Condens. Matter Phys.}}}}
	\textbf{\bibinfo{volume}{21}}, \bibinfo{pages}{395502}
	(\bibinfo{year}{2009}).
	
	\bibitem{nakata_large_2020}
	\bibinfo{author}{Nakata, A.} \emph{et~al.}
	\newblock \bibinfo{title}{{Large scale and linear scaling {DFT} with the
			{CONQUEST} code}}.
	\newblock \emph{\bibinfo{journal}{\emph{\emph{J. Chem. Phys.}}}}
	\textbf{\bibinfo{volume}{152}}, \bibinfo{pages}{164112}
	(\bibinfo{year}{2020}).
	
	\bibitem{wood_data-driven_2019}
	\bibinfo{author}{Wood, M.~A.}, \bibinfo{author}{Cusentino, M.~A.},
	\bibinfo{author}{Wirth, B.~D.} \& \bibinfo{author}{Thompson, A.~P.}
	\newblock \bibinfo{title}{{Data-driven material models for atomistic
			simulation}}.
	\newblock \emph{\bibinfo{journal}{\emph{\emph{Phys. Rev. B}}}}
	\textbf{\bibinfo{volume}{99}}, \bibinfo{pages}{184305}
	(\bibinfo{year}{2019}).
	
	\bibitem{wilkins_accurate_2019}
	\bibinfo{author}{Wilkins, D.~M.} \emph{et~al.}
	\newblock \bibinfo{title}{{Accurate Molecular Polarizabilities with Coupled
			Cluster Theory and Machine Learning}}.
	\newblock \emph{\bibinfo{journal}{\emph{\emph{Proc. Natl. Acad. Sci. U.S.A.}}}}
	\textbf{\bibinfo{volume}{116}}, \bibinfo{pages}{3401--3406}
	(\bibinfo{year}{2019}).
	
	\bibitem{OVITO}
	\bibinfo{author}{Stukowski, A.}
	\newblock \bibinfo{title}{{Visualization and analysis of atomistic simulation
			data with {OVITO}{\textendash}the Open Visualization Tool}}.
	\newblock \emph{\bibinfo{journal}{\emph{\emph{Model. Simul. Mat. Sci. Eng.}}}}
	\textbf{\bibinfo{volume}{18}}, \bibinfo{pages}{015012}
	(\bibinfo{year}{2009}).
	
	\bibitem{vmd}
	\bibinfo{author}{Humphrey, W.}, \bibinfo{author}{Dalke, A.} \&
	\bibinfo{author}{Schulten, K.}
	\newblock \bibinfo{title}{{VMD: visual molecular dynamics}}.
	\newblock \emph{\bibinfo{journal}{\emph{\emph{J. Mol. Graph.}}}}
	\textbf{\bibinfo{volume}{14}}, \bibinfo{pages}{33--38}
	(\bibinfo{year}{1996}).
	
	\bibitem{kresse_ab_1993}
	\bibinfo{author}{Kresse, G.} \& \bibinfo{author}{Hafner, J.}
	\newblock \bibinfo{title}{{Ab initio molecular Dynamics for Liquid Metals}}.
	\newblock \emph{\bibinfo{journal}{\emph{\emph{Phys. Rev. B}}}}
	\textbf{\bibinfo{volume}{47}}, \bibinfo{pages}{558--561}
	(\bibinfo{year}{1993}).
	
	\bibitem{daw_embedded-atom_1984}
	\bibinfo{author}{Daw, M.~S.} \& \bibinfo{author}{Baskes, M.~I.}
	\newblock \bibinfo{title}{{Embedded-Atom Method: Derivation and Application to
			Impurities, Surfaces, and Other Defects in Metals}}.
	\newblock \emph{\bibinfo{journal}{\emph{\emph{Phys. Rev. B}}}}
	\textbf{\bibinfo{volume}{29}}, \bibinfo{pages}{6443--6453}
	(\bibinfo{year}{1984}).
	
	\bibitem{BeEAM}
	\bibinfo{author}{Agrawal, A.}, \bibinfo{author}{Mishra, R.},
	\bibinfo{author}{Ward, L.}, \bibinfo{author}{Flores, K.~M.} \&
	\bibinfo{author}{Windl, W.}
	\newblock \bibinfo{title}{An embedded atom method potential of beryllium}.
	\newblock \emph{\bibinfo{journal}{\emph{\emph{Model. Simul. Mat. Sci. Eng.}}}}
	\textbf{\bibinfo{volume}{21}}, \bibinfo{pages}{085001}
	(\bibinfo{year}{2013}).
	
	\bibitem{mermin_thermal_1965}
	\bibinfo{author}{Mermin, N.~D.}
	\newblock \bibinfo{title}{{Thermal Properties of the Inhomogeneous Electron
			Gas}}.
	\newblock \emph{\bibinfo{journal}{\emph{\emph{Phys. Rev.}}}}
	\textbf{\bibinfo{volume}{137}}, \bibinfo{pages}{A1441--A1443}
	(\bibinfo{year}{1965}).
	
	\bibitem{born_zur_1927}
	\bibinfo{author}{Born, M.} \& \bibinfo{author}{Oppenheimer, R.}
	\newblock \bibinfo{title}{{Zur Quantentheorie der Molekeln}}.
	\newblock \emph{\bibinfo{journal}{\emph{\emph{Ann. Phys.}}}}
	\textbf{\bibinfo{volume}{389}}, \bibinfo{pages}{457--484}
	(\bibinfo{year}{1927}).
	
	\bibitem{toda1983statistical}
	\bibinfo{author}{Toda, M.}, \bibinfo{author}{Kubo, R.}, \bibinfo{author}{Kubo,
		R.}, \bibinfo{author}{Sait{\=o}, N.} \& \bibinfo{author}{Hashitsume, N.}
	\newblock \emph{\emph{\bibinfo{title}{{\emph{Statistical Physics: Equilibrium
						statistical mechanics}}}}}.
	\newblock Solid-State Sciences Series (\bibinfo{publisher}{{Springer Berlin}},
	\bibinfo{address}{Heidelberg}, \bibinfo{year}{1983}).
	
	\bibitem{ceperley_ground_1980}
	\bibinfo{author}{Ceperley, D.~M.} \& \bibinfo{author}{Alder, B.~J.}
	\newblock \bibinfo{title}{{Ground State of the Electron Gas by a Stochastic
			Method}}.
	\newblock \emph{\bibinfo{journal}{\emph{\emph{Phys. Rev. Letters}}}}
	\textbf{\bibinfo{volume}{45}}, \bibinfo{pages}{566--569}
	(\bibinfo{year}{1980}).
	
	\bibitem{perdew_accurate_1986}
	\bibinfo{author}{Perdew, J.~P.} \& \bibinfo{author}{Yue, W.}
	\newblock \bibinfo{title}{{Accurate and Simple Density Functional for the
			Electronic Exchange Energy: Generalized Gradient Approximation}}.
	\newblock \emph{\bibinfo{journal}{\emph{\emph{Phys. Rev. B}}}}
	\textbf{\bibinfo{volume}{33}}, \bibinfo{pages}{8800--8802}
	(\bibinfo{year}{1986}).
	
	\bibitem{perdew_accurate_1992}
	\bibinfo{author}{Perdew, J.~P.} \& \bibinfo{author}{Wang, Y.}
	\newblock \bibinfo{title}{{Accurate and Simple Analytic Representation of the
			Electron-Gas Correlation Energy}}.
	\newblock \emph{\bibinfo{journal}{\emph{\emph{Phys. Rev. B}}}}
	\textbf{\bibinfo{volume}{45}}, \bibinfo{pages}{13244--13249}
	(\bibinfo{year}{1992}).
	
	\bibitem{perdew_generalized_1996}
	\bibinfo{author}{Perdew, J.~P.}, \bibinfo{author}{Burke, K.} \&
	\bibinfo{author}{Ernzerhof, M.}
	\newblock \bibinfo{title}{{Generalized Gradient Approximation Made Simple}}.
	\newblock \emph{\bibinfo{journal}{\emph{\emph{Phys. Rev. Lett.}}}}
	\textbf{\bibinfo{volume}{77}}, \bibinfo{pages}{3865--3868}
	(\bibinfo{year}{1996}).
	
	\bibitem{sun_strongly_2015}
	\bibinfo{author}{Sun, J.}, \bibinfo{author}{Ruzsinszky, A.} \&
	\bibinfo{author}{Perdew, J.~P.}
	\newblock \bibinfo{title}{{Strongly Constrained and Appropriately Normed
			Semilocal Density Functional}}.
	\newblock \emph{\bibinfo{journal}{\emph{\emph{Phys. Rev. Letters}}}}
	\textbf{\bibinfo{volume}{115}}, \bibinfo{pages}{036402}
	(\bibinfo{year}{2015}).
	
	\bibitem{brown_2013_exchange}
	\bibinfo{author}{Brown, E.~W.}, \bibinfo{author}{DuBois, J.~L.},
	\bibinfo{author}{Holzmann, M.} \& \bibinfo{author}{Ceperley, D.~M.}
	\newblock \bibinfo{title}{{Exchange-correlation energy for the
			three-dimensional homogeneous electron gas at arbitrary temperature}}.
	\newblock \emph{\bibinfo{journal}{\emph{\emph{Phys. Rev. B}}}}
	\textbf{\bibinfo{volume}{88}}, \bibinfo{pages}{081102}
	(\bibinfo{year}{2013}).
	
	\bibitem{karasiev_nonempirical_2013}
	\bibinfo{author}{Karasiev, V.~V.}, \bibinfo{author}{Chakraborty, D.},
	\bibinfo{author}{Shukruto, O.~A.} \& \bibinfo{author}{Trickey, S.~B.}
	\newblock \bibinfo{title}{{Nonempirical Generalized Gradient Approximation
			Free-Energy Functional for Orbital-Free Simulations}}.
	\newblock \emph{\bibinfo{journal}{\emph{\emph{Phys. Rev. B}}}}
	\textbf{\bibinfo{volume}{88}}, \bibinfo{pages}{161108}
	(\bibinfo{year}{2013}).
	
	\bibitem{karasiev_2014_accurate}
	\bibinfo{author}{Karasiev, V.~V.}, \bibinfo{author}{Sjostrom, T.},
	\bibinfo{author}{Dufty, J.} \& \bibinfo{author}{Trickey, S.~B.}
	\newblock \bibinfo{title}{{Accurate Homogeneous Electron Gas
			Exchange-Correlation Free Energy for Local Spin-Density Calculations}}.
	\newblock \emph{\bibinfo{journal}{\emph{\emph{Phys. Rev. Lett.}}}}
	\textbf{\bibinfo{volume}{112}}, \bibinfo{pages}{076403}
	(\bibinfo{year}{2014}).
	
	\bibitem{groth_2017_ab}
	\bibinfo{author}{Groth, S.} \emph{et~al.}
	\newblock \bibinfo{title}{{Ab initio Exchange-Correlation Free Energy of the
			Uniform Electron Gas at Warm Dense Matter Conditions}}.
	\newblock \emph{\bibinfo{journal}{\emph{\emph{Phys. Rev. Lett.}}}}
	\textbf{\bibinfo{volume}{119}}, \bibinfo{pages}{135001}
	(\bibinfo{year}{2017}).
	
	\bibitem{MDreview}
	\bibinfo{author}{Iftimie, R.}, \bibinfo{author}{Minary, P.} \&
	\bibinfo{author}{Tuckerman, M.~E.}
	\newblock \bibinfo{title}{{Ab initio molecular dynamics: Concepts, recent
			developments, and future trends}}.
	\newblock \emph{\bibinfo{journal}{\emph{\emph{Proc. Natl. Acad. Sci. U.S.A.}}}}
	\textbf{\bibinfo{volume}{102}}, \bibinfo{pages}{6654--6659}
	(\bibinfo{year}{2005}).
	
	\bibitem{FiSh2022}
	\bibinfo{author}{Fiedler, L.}, \bibinfo{author}{Shah, K.},
	\bibinfo{author}{Bussmann, M.} \& \bibinfo{author}{Cangi, A.}
	\newblock \bibinfo{title}{{Deep dive into machine learning density functional
			theory for materials science and chemistry}}.
	\newblock \emph{\bibinfo{journal}{\emph{\emph{Phys. Rev. Materials}}}}
	\textbf{\bibinfo{volume}{6}}, \bibinfo{pages}{040301} (\bibinfo{year}{2022}).
	
	\bibitem{chandrasekaran_solving_2019}
	\bibinfo{author}{Chandrasekaran, A.} \emph{et~al.}
	\newblock \bibinfo{title}{{Solving the Electronic Structure Problem with
			Machine Learning}}.
	\newblock \emph{\bibinfo{journal}{\emph{\emph{npj Comput. Mater.}}}}
	\textbf{\bibinfo{volume}{5}}, \bibinfo{pages}{22} (\bibinfo{year}{2019}).
	
	\bibitem{thompson_spectral_2015}
	\bibinfo{author}{Thompson, A.~P.}, \bibinfo{author}{Swiler, L.~P.},
	\bibinfo{author}{Trott, C.~R.}, \bibinfo{author}{Foiles, S.~M.} \&
	\bibinfo{author}{Tucker, G.~J.}
	\newblock \bibinfo{title}{{Spectral {Neighbor} {Analysis} {Method} for
			{Automated} {Generation} of {Quantum}-{Accurate} {Interatomic}
			{Potentials}}}.
	\newblock \emph{\bibinfo{journal}{\emph{\emph{J. Comput. Phys.}}}}
	\textbf{\bibinfo{volume}{285}}, \bibinfo{pages}{316--330}
	(\bibinfo{year}{2015}).
	
	\bibitem{fiedler2022electronic}
	\bibinfo{author}{Fiedler, L.} \emph{et~al.}
	\newblock \bibinfo{title}{{Training-free hyperparameter optimization of neural
			networks for electronic structures in matter}}.
	\newblock \emph{\bibinfo{journal}{\emph{\emph{ Mach. Learn.: Sci. Technol.}}}}
	\textbf{\bibinfo{volume}{3}}, \bibinfo{pages}{045008} (\bibinfo{year}{2022}).
	
	\bibitem{hornik_approximation_1991}
	\bibinfo{author}{Hornik, K.}
	\newblock \bibinfo{title}{{Approximation Capabilities of Multilayer Feedforward
			Networks}}.
	\newblock \emph{\bibinfo{journal}{\emph{\emph{ Neural Netw.}}}}
	\textbf{\bibinfo{volume}{4}}, \bibinfo{pages}{251--257}
	(\bibinfo{year}{1991}).
	
	\bibitem{minsky_perceptrons_1987}
	\bibinfo{author}{Minsky, M.} \& \bibinfo{author}{Papert, S.~A.}
	\newblock \emph{\emph{\bibinfo{title}{\emph{{Perceptrons. {{An Introduction}}
						to {{Computational Geometry}}}}}}} (\bibinfo{publisher}{{MIT Press}},
	\bibinfo{address}{{Cambridge, MA}}, \bibinfo{year}{2017}).
	
	\bibitem{rosenblatt_perceptron_1957}
	\bibinfo{author}{Rosenblatt, F.}
	\newblock \emph{\emph{\bibinfo{title}{\emph{{The {{Perceptron}}: A
						{{Perceiving}} and {{Recognizing Automaton}} ({{Project PARA}}).}}}}}
	(\bibinfo{publisher}{{Cornell Aeronautical Laboratory}},
	\bibinfo{address}{{Buffalo, NY}}, \bibinfo{year}{1957}).
	
	\bibitem{rumelhart1986learning}
	\bibinfo{author}{Rumelhart, D.~E.}, \bibinfo{author}{Hinton, G.~E.} \&
	\bibinfo{author}{Williams, R.~J.}
	\newblock \bibinfo{title}{{Learning representations by back-propagating
			errors}}.
	\newblock \emph{\bibinfo{journal}{\emph{\emph{Nature}}}}
	\textbf{\bibinfo{volume}{323}}, \bibinfo{pages}{533--536}
	(\bibinfo{year}{1986}).
	
	\bibitem{deeplearningbook}
	\bibinfo{author}{Goodfellow, I.}, \bibinfo{author}{Bengio, Y.} \&
	\bibinfo{author}{Courville, A.}
	\newblock \emph{\emph{\bibinfo{title}{{\emph{Deep Learning}}}}}
	(\bibinfo{publisher}{MIT Press}, \bibinfo{address}{{Cambridge, MA}},
	\bibinfo{year}{2016}).
	
	\bibitem{rdf}
	\bibinfo{author}{Allen, M.~P.} \& \bibinfo{author}{Tildesley, D.~J.}
	\newblock \emph{\emph{\bibinfo{title}{\emph{{Computer simulation of
						liquids}}}}} (\bibinfo{publisher}{Oxford University Press},
	\bibinfo{address}{Oxford}, \bibinfo{year}{1989}).
	
	\bibitem{metropolis1953}
	\bibinfo{author}{Metropolis, N.}, \bibinfo{author}{Rosenbluth, A.~W.},
	\bibinfo{author}{Rosenbluth, M.~N.}, \bibinfo{author}{Teller, A.~H.} \&
	\bibinfo{author}{Teller, E.}
	\newblock \bibinfo{title}{{Equation of State Calculations by Fast Computing
			Machines}}.
	\newblock \emph{\bibinfo{journal}{\emph{\emph{J. Chem. Phys.}}}}
	\textbf{\bibinfo{volume}{21}}, \bibinfo{pages}{1087--1092}
	(\bibinfo{year}{1953}).
	
	\bibitem{kresse_efficient_1996}
	\bibinfo{author}{Kresse, G.} \& \bibinfo{author}{Furthm{\"u}ller, J.}
	\newblock \bibinfo{title}{{Efficient Iterative Schemes Forab Initiototal-Energy
			Calculations Using a Plane-Wave Basis Set}}.
	\newblock \emph{\bibinfo{journal}{\emph{\emph{Phys. Rev. B}}}}
	\textbf{\bibinfo{volume}{54}}, \bibinfo{pages}{11169--11186}
	(\bibinfo{year}{1996}).
	
	\bibitem{kresse_efficiency_1996}
	\bibinfo{author}{Kresse, G.} \& \bibinfo{author}{Furthm{\"u}ller, J.}
	\newblock \bibinfo{title}{{Efficiency of Ab-Initio Total Energy Calculations
			for Metals and Semiconductors Using a Plane-Wave Basis Set}}.
	\newblock \emph{\bibinfo{journal}{\emph{\emph{Comput. Mater. Sci.}}}}
	\textbf{\bibinfo{volume}{6}}, \bibinfo{pages}{15--50} (\bibinfo{year}{1996}).
	
	\bibitem{QEPseudopotentials}
	\bibinfo{author}{{Dal Corso}, A.}
	\newblock \bibinfo{title}{{Pseudopotentials periodic table: From H to Pu}}.
	\newblock \emph{\bibinfo{journal}{\emph{\emph{Comput. Mater. Sci.}}}}
	\textbf{\bibinfo{volume}{95}}, \bibinfo{pages}{337--350}
	(\bibinfo{year}{2014}).
	
	\bibitem{blochl_projector_1994}
	\bibinfo{author}{Bl{\"o}chl, P.~E.}
	\newblock \bibinfo{title}{{Projector augmented-wave method}}.
	\newblock \emph{\bibinfo{journal}{\emph{\emph{Phys. Rev. B}}}}
	\textbf{\bibinfo{volume}{50}}, \bibinfo{pages}{17953--17979}
	(\bibinfo{year}{1994}).
	
	\bibitem{kresse_ultrasoft_1999}
	\bibinfo{author}{Kresse, G.} \& \bibinfo{author}{Joubert, D.}
	\newblock \bibinfo{title}{{From ultrasoft pseudopotentials to the projector
			augmented-wave method}}.
	\newblock \emph{\bibinfo{journal}{\emph{\emph{Phys. Rev. B}}}}
	\textbf{\bibinfo{volume}{59}}, \bibinfo{pages}{1758--1775}
	(\bibinfo{year}{1999}).
	
	\bibitem{Orbital_Free_DFTMD}
	\bibinfo{author}{Fiedler, L.} \emph{et~al.}
	\newblock \bibinfo{title}{{Accelerating equilibration in first-principles
			molecular dynamics with orbital-free density functional theory}}.
	\newblock \emph{\bibinfo{journal}{\emph{\emph{Phys. Rev. Research}}}}
	\textbf{\bibinfo{volume}{4}}, \bibinfo{pages}{043033} (\bibinfo{year}{2022}).
	
	\bibitem{monkhorst_special_1976}
	\bibinfo{author}{Monkhorst, H.~J.} \& \bibinfo{author}{Pack, J.~D.}
	\newblock \bibinfo{title}{{Special points for {Brillouin}-zone integrations}}.
	\newblock \emph{\bibinfo{journal}{\emph{\emph{Phys. Rev. B}}}}
	\textbf{\bibinfo{volume}{13}}, \bibinfo{pages}{5188--5192}
	(\bibinfo{year}{1976}).
	
	\bibitem{fiedler_ldossnap_2022}
	\bibinfo{author}{Fiedler, L.} \& \bibinfo{author}{Cangi, A.}
	\newblock \bibinfo{title}{Data set '{LDOS}/{SNAP} data for {MALA}: Beryllium at
		{298K}'}.
	\newblock \emph{\bibinfo{journal}{\emph{RODARE,
				https://doi.org/10.14278/rodare.1834}}}  (\bibinfo{year}{2022}).
	
	\bibitem{sizetransferscripts}
	\bibinfo{author}{Fiedler, L.} \emph{et~al.}
	\newblock \bibinfo{title}{{Data set 'Scripts and Models for "Predicting
			electronic structures at any length scale with machine learning"'}}.
	\newblock \emph{\bibinfo{journal}{\emph{RODARE,
				https://doi.org/10.14278/rodare.1851}}}  (\bibinfo{year}{2022}).
	
\end{thebibliography}
\end{document}